\documentclass[journal,onecolumn,12pt]{IEEEtran}
\IEEEoverridecommandlockouts %showing the thanks at the left bottom of the first page
\usepackage{epsfig,latexsym}
\usepackage{float}
\usepackage{indentfirst}
\usepackage{amsmath}
\usepackage{amssymb}
\usepackage{times}
\usepackage{subfigure}
\usepackage{pifont}
\usepackage{psfrag}
\usepackage{cite}
\usepackage{url}
\usepackage{lastpage}
\usepackage{epstopdf}
\usepackage{bm}
\usepackage{color}
\usepackage{cases}
\usepackage{fancyhdr}
\usepackage[center]{caption2}
\usepackage{balance}
\usepackage{fancyhdr,lastpage}
\begin{document}
\title{Resource Allocation Strategies for Secure WPCN Multiantenna Multicasting Systems}
\author{Zheng Chu, Huan X. Nguyen,~\IEEEmembership{Senior Member,~IEEE}, Giuseppe Caire,~\IEEEmembership{Fellow,~IEEE}
%%         % <-this % stops a space
\thanks{Z. Chu and H. Nguyen are with the School of Science and Technology, Middlesex University, The Burroughs, London NW4 4BT, U.K. (Email: z.chu@mdx.ac.uk%h.nguyen@mdx.ac.uk
	)}% <-this % stops a space
\thanks{Giuseppe Caire is with the Communications and Information Theory Group, Technical University of Berlin, 10587 Berlin, Germany, and also with the Department of Electrical Engineering, University of Southern California, Los Angeles, CA 90089 USA. %(Email: caire@tu-berlin.de)
}
}
\maketitle
\thispagestyle{empty}
\vspace{-1em}
\begin{abstract}
This paper investigates a secure wireless-powered multiantenna multicasting system, where multiple power beacons (PBs) supply power to a transmitter in order to establish a reliable communication link with multiple legitimate users in the presence of multiple eavesdroppers. The transmitter has to harvest radio frequency (RF) energy from multiple PBs due to the shortage of embedded power supply before establishing its secure communication. We consider two different scenarios. In the first, the PBs and the transmitter belong to the same operator, where we formulate the resource allocation problem as the minimization of the total transmit power subject to the target secure rate constraint. The solution of this problem yields both the optimal power and energy transfer time allocation. Due to the non-convexity of this problem, we propose a two-level approach, where the inner level problem can be recast as a convex optimization framework via conic convex reformulation, while the outer level problem can be handled by using one-dimensional (1D) search. The second scenario considers the case where the transmitter and the PBs belong to different service suppliers. Hence, we formulate the resource allocation problem where we consider incentives for the PBs to assist the transmitter. This leads to the formulation of a \emph{Stackelberg} game for the secure wireless-powered multiantenna multicasting system. 
% Moreover, a \emph{\emph{Stackelberg}} game is formulated for this secure wireless-powered MISO multicasting system when incentives are considered for the PBs to assist the transmitter since the transmitter and the PBs may belong to different service suppliers. 
 The transmitter has to pay for the energy services from these multiple PBs in order to facilitate secure communications. In this game, the transmitter and the PB are modelled as leader and follower, respectively, in which both of them try to maximize their own utility function. The closed-form \emph{Stackelberg} equilibrium of the formulated game is then derived. Finally, numerical results are provided to validate our proposed schemes.
\end{abstract}

\IEEEpeerreviewmaketitle
\setlength{\baselineskip}{1\baselineskip}
\newtheorem{definition}{Definition}
\newtheorem{fact}{Fact}
\newtheorem{assumption}{Assumption}
\newtheorem{theorem}{Theorem}
\newtheorem{lemma}{Lemma}
\newtheorem{corollary}{Corollary}
\newtheorem{proposition}{Proposition}
\newtheorem{example}{Example}
\newtheorem{remark}{Remark}
\newtheorem{algorithm}{Algorithm}
\newcommand{\mv}[1]{\mbox{\boldmath{$ #1 $}}}

% \fi
\IEEEpeerreviewmaketitle
\section{Introduction}
 Wireless multicast media streaming is anticipated to be a significant component of the forthcoming 5G systems, motivated by the consumers' desire to take advantage of high quality multimedia wireless devices (e.g., 4k hand-held devices, 3D augmented reality)\cite{Tom_Luo_Multicasting_TSP_2006,Zhengzheng_Xiang_TWC_Multicasting_2013,Zhengzheng_Xiang_JSAC_Multicasting_2014}. Energy efficiency and security are major critical issues that must be addressed in the design of such systems.
 
Radio frequency (RF) energy harvesting and transfer techniques have recently been considered as a promising solution to the energy-constrained wireless networks \cite{Varshney_08,Shannon_Tesla_10,Zhang_Rui_SWIPT_TWC13}. 
%This innovation has greatly broadened the application of energy-constrained networks to extend limited battery lifetime.
As a recent application of RF energy harvesting and transfer techniques, wireless powered communication networks (WPCNs) have become a novel technology in wireless networking and attracted more and more attention \cite{Rui_Zhang_WPCN_CM_2016}. A ``\emph{harvest-then-transmit}'' protocol was proposed for WPCNs in \cite{Rui_Zhang_WPCN_TWC_2014}, where the wireless users harvest power from the RF signals broadcast by an access point in the downlink (DL), and then send information to the AP in the uplink (UL) by employing the harvested energy. Cooperative protocols for WPCNs were developed based on different models \cite{Rui_Zhang_GLOBECOM_2014_UC_WPCN,He_Chen_ITW_WPCN_Relay_2014,He_Chen_TSP_HTC_WPCN_2016}.
A different approach consists of deploying a dedicated wireless energy transfer (WET) network with multiple power beacons (PBs) to provide wireless charging services to the wireless terminals via the RF energy transfer technique \cite{Kaibin_Huang_CM_2015,Kaibin_Huang_TWC_2014_WPT}. %Moreover, cellular base stations (BSs) are underlaid by these PBs, which transfer the information to mobile devices by using the harvested energy. 
Since the PBs do not require any \emph{backhaul link}, the associated cost of PBs deployment is much lower, hence, it is feasible to deploy the PBs densely to guarantee network coverage for a wide range of mobile devices \cite{Caijun_Zhong_SPL_2015_WPR}.

Security in data transmission can be addressed either by traditional crypto methods, or more fundamentally, in terms of information theoretic secure rates. The  latter approach, commonly referred to as ``physical-layer security,'' was initially developed for the wiretap channel \cite{Wyner_J75,Korner_Info_Theory_J78}, i.e., a broadcast channel with one transmitter and two sets of receivers: legitimate users and eavesdroppers.  
%Physical-layer security, as an alternative approach to traditional cryptographic methods, has been developed by initially defining wiretap channel based on information theory principle to improve the security of wireless transmission in the physical layer \cite{Wyner_J75,Korner_Info_Theory_J78}. 
Multiantenna wiretap channels have been widely investigated in terms of secure rate region \cite{Shamai_Multiantenna_Wiretap_TIT_2009,Wornell_Info_Theory_J10,Wornell_Info_Theory1_J10,
Petropulu_Gaussian_MISO_Wiretap_TWC_2011,Swindlehurst_FR_MIMO_Wiretap_TSP_2013}. Some state-of-art techniques, such as artificial noise (AN) and cooperative jammer (CJ), have been designed for multiantenna transceivers, in order to introduce more interference at the eavesdroppers \cite{Ma_Sig_Process_J11,Ma_TSP_J13,Zheng_Secrecy_J15,Zheng_Stackelberg_game_EUSIPCO_2014,Zheng_WCL_2015,Zheng_Sec_TWC_2016}. In \cite{Ma_Sig_Process_J11}, 
%mathematical optimization techniques were employed in the MISO secure channels to solve the secrecy rate optimization problems (i.e., power minimization and secrecy rate maximization) in order to guarantee the reliable communications. 
rank-one solution properties were exploited with semidefinite programming (SDP) relaxation for secure transmit beamforming. AN-assisted transmit optimization has been presented in \cite{Ma_TSP_J13}, where the spatially selective AN embedded with secure transmit beamforming was designed to obtain the optimal power allocation. In \cite{Zheng_Secrecy_J15}, CJ from an external node is exploited in order to create interference at the eavesdroppers and achieve the desired target secure rate. However, it is not always possible to have an own CJ to improve the secrecy rates. Another option could be to employ a private CJ by paying a price for the jamming services. This strategy was investigated in  \cite{Zheng_Secrecy_J15,Zheng_Stackelberg_game_EUSIPCO_2014},  where a 
 CJ releases its jamming service depending interference caused to the eavesdropper, while the transmitter pays a certain amount to guarantee its secure communication. In this strategy, a \emph{Stackelberg} game can be formulated to obtain the optimal power allocation. In addition, cooperative cognitive radio (CR) combined with secure communications could also be modelled as a \emph{Stackelberg} game to determine the optimal resource allocations \cite{Nallan_TCOM_2015_CCR_Sec_Game}. In \cite{Zheng_WCL_2015,Zheng_Sec_TWC_2016}, the secrecy rate optimization problem was posed in terms of outage secrecy rates, due to the fact that the channels are not perfectly known and are subject to random fading with known statistics. 
%outage secrecy rate optimizations were formulated based on the practical assumption that the statistical channel uncertainties are modelled, where \emph{Bernstein-type} inequality can be employed to relax these formulated problems. 
 Physical-layer security techniques have also been recently developed in radio frequency identification (RFID). The design of RFID systems is a challenge due to the broadcast nature of backscatter communication, which is vulnerable to eavesdropping  \cite{Huiming_Wang_RFID_Backscatter_PLS_TWC_2016}. Simultaneous wireless information and power transfer (SWIPT) has emerged as one  of most promising approaches to provide power for communication devices. SWIPT has been considered in combination with physical-layer security in a number of recent works (e.g., \cite{Zhang_Rui_TSP_Sec_SWIPT_J14,Derrick_SWIPT_TWC_2014,Khandaker_TIFS_J15,Zheng_TVT_SWIPT_2015}). It is worth pointing out that the transmit power is constant in the above secure communication systems. However, the use of WET effectively makes the available transmit power a system variable in order to achieve secure communications. Thus, this research gap motivates the work in this paper.
 
We investigate a WPCN-assisted multiantenna secure multicasting system in which a multicast service provider (i.e., the transmitter) guarantees secure communication with legitimate users in the presence of multiple eavesdroppers by utilizing the harvested energy from the PBs. In particular, we consider two different but complementary scenarios. In the first scenario, the transmitter and the PBs are considered to belong to the same service provider. In the second, they belong to different service providers with different objectives. Accordingly, we formulate and solve the two different problems as follows:  
 \begin{enumerate}
 	\item \emph{Power minimization for WPCN-aided multiantenna secure multicasting system}: This problem is formulated for the first scenario, where there is no energy trading between the transmitter and the PBs. They cooperate to minimize the transmit power for a given target secure multicast rate. The optimization problem is not jointly convex in terms of PBs' transmit power, the fraction of time dedicated to WET, and the transmit beamforming vector. To circumvent this non-convexity issue, we consider a two-level approach. The outer level can be recast as a single-variable optimization problem with respect to the energy time allocation, in which the optimal solution can be achieved via numerical search, while the inner level remains a non-convex power minimization problem. There are different approaches for the inner level problem. In \cite{Ma_Sig_Process_J11}, semidefinite programming (SDP) relaxation was employed to solve the power minimization for the case of single legitimate user only. However, this approach will not guarantee a rank-1 solution when extending to the case of multiple users. Therefore, a rank-1 feasible solution for the original beamforming problem must be obtained from the solution of the relaxed problem, with no optimality guarantee  (see, e.g., \cite{Qiang_Li_ICC_Sec_Multicasting_2011,Cumanan_JSTSP_2016_Game})
%(i.e., the SDP relaxed solution may \textit{not} be rank-one for the secure MISO multicasting muti-user system) such that it is now well documented that the SDP relaxed solution should be necessarily handled via a rank-reduction algorithm (see, e.g., \cite{Qiang_Li_ICC_Sec_Multicasting_2011}). 
Differently, we first propose a novel reformulation based on matrix transformations and convex conic optimization techniques, yielding a second-order cone programming (SOCP) solution which is optimal when the SDP relaxed solution satisfies the rank-one condition. Then, we propose a successive convex approximation (SCA) based SOCP scheme, which is performed iteratively to obtain the optimal transmit beamformer directly for any general case. Numerical results confirm that our proposed SCA based SOCP scheme outperforms the SDP relaxed scheme, and the proposed SOCP scheme has more computationally efficient than the SDP relaxed scheme that uses Guided Randomization method.
% 	our work focuses entirely on multiple users for multicasting system and our SOCP reformulation can directly yield the optimal beamforming vector.
% 	 whereas the inner level can be handled by a novel reformulation based on matrix transformations and convex conic optimization techniques, yielding a second-order cone programming (SOCP).
% 	The inner level optimization problem can be relaxed as a second order cone programming (SOCP) problem. 
 	%The result in \cite{Ulukus_ISIT_2007_MISO_Sec} cannot be extended to our paper because its optimal beamforming vector is achieved based on the assumption that the transmit power budget is available, whereas our work jointly designs the optimal power allocation and transmit beamforming vector.
% 	where we derive a closed-form optimal solution based on the dual problem and Karush-Kuhn-Tucker (KKT) conditions. 
 	\item \emph{Game theory based WPCN-aided multiantenna secure multicasting system}: In this problem, we investigate the case where the transmitter and the PBs belong to different service operators, both of which want to maximize their own benefit. Thus, an energy price must be paid by the transmitter in order to induce the PBs to provide enough WET to guarantee secure communications. 
 %To the best of the authors' knowledge, there has been no existing works that have investigated this energy interaction to guarantee secure communications. 
 In particular, we develop an energy trading framework for wireless powered secure multiantenna multicasting systems, where the strategic behavior of the transmitter and the PBs is modeled as a \emph{Stackelberg} game. The transmitter acts a leader that buys energy from the PBs to achieve a desired secure multicast rate to the users. The transmitter optimizes the energy price and energy transfer time to maximize its utility function, defined as the weighted difference between revenues (proportional to the achieved secure rate) and costs of the purchased energy. The PBs are the followers, that determine their optimal transmit powers based on the energy price released from the transmitter to
maximize their own profits, defined as the difference between the payment received from the transmitter and its energy cost. We obtain a closed-form solution for the Stackelberg equilibrium, in which both the PBs and the transmitter come to an agreement on the energy price, transmit power and energy time allocation. 	
 \end{enumerate}

\textbf{\textit{Related works:}}
Considering the first scenario, unlike [28] where an intermediate self-sustainable relay was employed to enable cooperation between a WET network and a wireless information transfer (WIT) network to guarantee secure communications subject to outage probability constraints, our work is to exploit direct energy interactions between the multicasting service provider and the PBs to facilitate secure communications of the legitimate users. Moreover, we also consider a special case with a single legitimate user and a single eavesdropper. In \cite{Ulukus_ISIT_2007_MISO_Sec}, the authors aimed to maximize the secrecy rate subject to the transmit power budget, and the optimal beamforming vector was designed based on Rayleigh quotient approach; whereas our work for this special case presents the closed-form solution of the joint design of the optimal time allocation and the optimal transmit beamforming vector to achieve the target of minimizing the transmitter's transmit power, via the dual problem and the Karush-Kuhn-Tucker (KKT) condition.

 The rest of the paper is organized as follows. Section \ref{section System model} presents our system model. Section \ref{section Power Efficient for WPCN MISO Secure System} solves the power minimization problem for the secure WPCN multiantenna multicasting system, whereas the game theory based secure WPCN multiantenna multicasting system is investigated in Section \ref{section Game_theory_based_secure_WPCN_MISO_multicasting_system}. Section \ref{section numerical_results} provides simulation results to validate the theoretical derivations. Finally, Section \ref{section Conclusion} concludes the paper.
 
 \subsection{Notations}
 \indent We use the upper case boldface letters for matrices and lower case boldface letters for vectors. $(\cdot)^{T}$ and $(\cdot)^{H}$ denote the transpose and conjugate transpose respectively. Tr$(\cdot)$ and $\mathbb{E}\{\cdot\}$ stand for trace of a matrix and the statistical expectation for random variables. $ \varrho_{max}(*) $ represents the maximum eigenvalue, whereas $ v_{\max}(*) $ denotes the eigenvector associated with the maximum eigenvalue. $\mathbf{A}\succeq 0$ indicates that $\mathbf{A}$ is a positive semidefinite matrix. $ \|*\| $ denotes the Euclidean norm of a vector. $\mathbf{I}$ and $(\cdot)^{-1}$ denote the identity matrix  with appropriate size and the inverse of a matrix respectively. 
% $ \|\cdot\| $ represents the Euclidean norm of a matrix. $\Re\{\cdot\}$ stands for the real part of a complex number, whereas 
% $|\mathbf{A}|$ denotes the determinant of $\mathbf{A}$. 
 $[x]^{+}$ represents $\max\{x,0\}$. The notation $ \succeq_{K_{n}} $ denotes the following generalized inequality: 
 	\begin{align}
 	\left[\begin{array}{cc}
 	a \\ \mathbf{b}
 	\end{array}
 	\right] \succeq_{K_{n}} \mathbf{0} \Leftrightarrow \|\mathbf{b}\| \leq a, \nonumber
 	\end{align} 
 where $ \mathbf{b} \in \mathbb{C}^{n-1} $ and $ K_{n} \subseteq \mathbb{R}^{n} $ is called a \emph{proper cone}.
\section{System Model}\label{section System model}
\begin{figure}[!htbp]
	\centering
	\includegraphics[scale = 1]{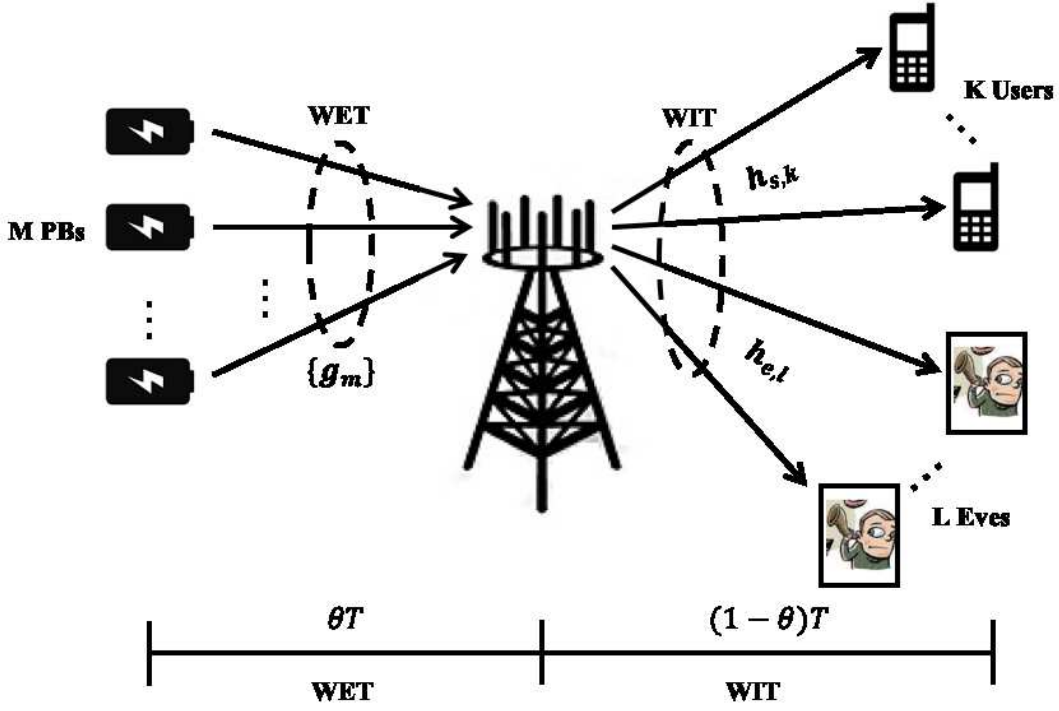}
	\caption{WPCN for multiantenna secure multicasting system.}
	\label{fig:WPCN_MISO_Secrecy}
\end{figure}

In this section, we consider the secure wireless powered multiantenna multicasting system as shown in Fig. \ref{fig:WPCN_MISO_Secrecy}, where a transmitter broadcasts the same information to all legitimate users in the presence of multiple eavesdroppers. Due to energy limitation at the transmitter, it is assumed that there is not enough power supply for information transfer, and thus the transmitter must harvest energy from the PBs. This system consists $ M $ single antenna PBs, one multiantenna transmitter equipped with $ N_{T} $ transmit antennas, $ K $ single antenna legitimate users and $ L $ single antenna eavesdroppers. This secrecy model has some potential applications, such as on-demand video broadcasting, wireless sensor networks and device to device (D2D) communication systems.
 In our paper, a \emph{harvest-then-transmit} protocol is considered. Specifically, time is divided in periods of duration $T$. Each period is split into a WET phase of duration $\theta T,$ and a WIT phase of duration $(1 - \theta) T,$ where $\theta \in (0,1)$ is a system parameter that must be optimized. 
 Let $ \mathbf{h}_{s,k} \in \mathbb{C}^{N_{T}\times 1} $ denote the channel coefficients between the transmitter and the $ k $-th legitimate user, while $  \mathbf{h}_{e,l} \in \mathbb{C}^{N_{T} \times 1} $ denotes the channel coefficients between the transmitter and the $ l $-th eavesdropper. 
 Also, $ \mathbf{g}_{m} \in \mathbb{C}^{1\times N_{T}} $ denotes the channel coefficients between the $ m $-th PB and the transmitter. 
First, each PB transfers the energy to the transmitter, the harvested energy during the WET phase of $ \theta T $ at the transmitter can be written as 
\begin{align}\label{eq:E_BS}
E_{B} = \xi \sum_{m=1}^{M}p_{m}\|\mathbf{g}_{m}\|^{2} \theta T,
\end{align}
where $ p_{m} $ denotes the transmit power of the $ m $-th PB, and $ 0 < \xi \leq 1  $ is the efficiency for converting the harvested energy to the electrical energy to be stored, which is assumed to be $ \xi = 1 $ in this paper. 
%Note that the thermal noise of the total harvested energy is ignored in \eqref{eq:E_BS} without loss of generality.
During the WIT phase of $ (1-\theta)T $, the received signal at the $ k $-th legitimate user and the $ l $-th  eavesdropper are given by  
 \begin{align}
 \mathbf{y}_{s,k} & = \sqrt{\frac{E_{B}}{(1-\theta)T}}\mathbf{h}_{s,k}^{H}\mathbf{v}s + n_{s,k}, ~k = 1, ..., K,\nonumber\\
 \mathbf{y}_{e,l} & = \sqrt{\frac{E_{B}}{(1-\theta)T}}\mathbf{h}_{e,l}^{H}\mathbf{v}s + n_{e,l}, ~l = 1,...,L,\nonumber
 \end{align}
 where $s$ denotes the Gaussian distributed transmit signal with unit norm, $ \mathbf{v} \in \mathbb{C}^{N_{T} \times 1} $ is the normalized transmit beamformer with $ \mathbb{E}\{\|\mathbf{v}\|^{2}\} = 1 $, $ n_{s,k} $ and $ n_{e,l} $ are additive white Gaussian noises (AWGNs) at the $ k $-th legitimate user and the $ l $-th eavesdropper with variance $ \sigma_{s}^{2} $ and $ \sigma_{e}^{2} $. 
Hence, the channel capacity of the $ k $-th legitimate user and the $ l $-th eavesdropper can be expressed as \cite{Xiaoming_Chen_CL_2016_Sec_WPCN}
\begin{align}
R_{s,k} = (1-\theta)\log \left( 1+ \frac{\theta \sum_{m=1}^{M}p_{m}\|\mathbf{g}_{m}\|^{2} |\mathbf{h}_{s,k}^{H}\mathbf{v}|^{2}}{(1-\theta)\sigma_{s}^{2}} \right), ~\forall k,
\end{align} 
and
\begin{align}
R_{e,l} = (1-\theta)\log \left( 1+ \frac{\theta \sum_{m=1}^{M}p_{m}\|\mathbf{g}_{m}\|^{2} |\mathbf{h}_{e,l}^{H}\mathbf{v}|^{2}}{(1-\theta)\sigma_{e}^{2}} \right),~\forall l,
\end{align} 
respectively. 
For this secure multicasting system, we have the following definition:
\begin{definition}
	Multicast secrecy rate of a multicasting system with $ K $ users is defined as \cite{Cumanan_JSTSP_2016_Game}
	\begin{align}
	R_{K} = \min_{k \in [1,K]} [R_{s,k} - \max_{l \in [1,L]} R_{e,l}]^{+}.
	\end{align}
\end{definition}
%According to the definition \cite{Wyner_J75}, the achievable secrecy rate at the $ k $-th legitimate user can be written as 
%\begin{align}
%R_{k} = [R_{s,k} - \max_{1\leq l \leq L} R_{e,l}]^{+},~\forall k,l.
%\end{align}
In the following, we develop two resource allocation schemes depending on whether the transmitter and the PBs belong to the same service supplier. 
\section{Power Minimization for WPCN Multiantenna Secure Multicasting System}\label{section Power Efficient for WPCN MISO Secure System}
In this section, we consider the scenario that the transmitter and the PBs belong to the same service provider in a multicasting network. 
%This means that they will work together to achieve a common target of maximizing the network's energy efficiency while maintaining the required level of information security. 
We will formulate a power minimization problem where the total transmit power at the transmitter is minimized to satisfy the target secrecy rate for all the legitimate users in the presence of multiple eavesdroppers by using the harvested energy from the PBs. We assume that the channel state information (CSI) between the transmitter and $ k $-th user as well as $ l $-th eavesdropper (i.e., $ \mathbf{h}_{s,k},~\forall k $ and $ \mathbf{h}_{e,l},~\forall l $) is available at the transmitter. This can be achieved through different methods such as the local oscillator power leakage from the eavesdropper receivers' RF frontend \cite{Mukherjee_ICASSP_2012} or even the CSI feedback method \cite{Geraci_TCOM_2012}. For example, in a video broadcasting system there may be legitimate users that are entitled to receive the content and other users who have not subscribed to this content, but are still part of the system. These users obey the basic physical-layer protocol rules, which includes feeding back CSI to enable beamforming. Hence, in this case, it is practical to assume that the CSI of the eavesdroppers is known at the transmitter.

\subsection{Power Minimization}
The minimization of the total transmit power\footnote{The objective function in \eqref{eq:Power_efficient_ori} consists of three variables: the transmit power of the PBs (i.e., $ p_{m} $, $ \forall m $), energy transfer time allocation (i.e., $ \theta $) and normalized transmit beamforming vector (i.e., $ \mathbf{v} $).} subject to a given secure multicast rate constraint can be written as:
\begin{align}\label{eq:Power_efficient_ori}
%\min_{\mathbf{v},\theta} &~ \frac{\frac{\theta}{1-\theta} \|\mathbf{v}\|^{2}}{ E_{BS} } = \frac{\frac{\theta}{1-\theta}(\sum_{m=1}^{M} p_{m}\| \mathbf{g}_{m} \|^{2}) \|\mathbf{v}\|^{2}}{\theta  \xi \sum_{m=1}^{M} p_{m}\|\mathbf{g}_{m}\|^{2}}  \nonumber\\
%s.t. &~ \min_{k} R_{k} \geq \bar{R}, ~ 0 \leq \theta \leq 1,~ \|\mathbf{v}\|^{2} = 1.
\min_{p_{m},\mathbf{v},\theta}&~ \frac{\sum_{m=1}^{M}p_{m}\|\mathbf{g}_{m}\|^{2} \theta }{(1-\theta)} \|\mathbf{v}\|^{2}, \nonumber\\
s.t. &~  R_{K} \geq \bar{R}, 
~0 \leq p_{m} \leq P,
~ 0 < \theta < 1,%~ \|\mathbf{v}\|^{2} = 1,
\end{align} 
where 
$P$ is the maximum power constraint at each PB, and 
$ \bar{R} $ denotes the target secure multicast rate. The problem \eqref{eq:Power_efficient_ori} is not convex in terms of $ \mathbf{v} $ and $ \theta $, and cannot be solved directly. It is obvious that from problem \eqref{eq:Power_efficient_ori}, the optimal solution of $p_m$ is $p_m = P,$ which %means that each PB has a full power policy before carrying out energy transfer to the transmitter.
is the maximum transmit power at each PB. 
Thus, the problem \eqref{eq:Power_efficient_ori} reduces to
\begin{align}\label{eq:Power_efficient_ori_rewritten}
\min_{\mathbf{w},\theta}&~ \frac{\sum_{m=1}^{M}\|\mathbf{g}_{m}\|^{2} \theta }{(1-\theta)} \|\mathbf{w}\|^{2}, \nonumber\\
s.t. &~  R_{K}(\theta,\mathbf{w}) \geq \bar{R}, ~ 0 < \theta < 1,
\end{align} 
where $ R_{K}(\theta,\mathbf{w}) $ can be written as 
\begin{align}
R_{K}(\theta, \mathbf{w}) = \min_{k \in [1,K]} (1-\theta)\bigg[ \log \bigg( 1+ \frac{\theta \sum_{m=1}^{M}\|\mathbf{g}_{m}\|^{2} |\mathbf{h}_{s,k}^{H}\mathbf{w}|^{2}}{(1-\theta)\sigma_{s}^{2}} \bigg) - \max_{l \in [1,L]}\log \bigg( 1+ \frac{\theta \sum_{m=1}^{M}\|\mathbf{g}_{m}\|^{2} |\mathbf{h}_{e,l}^{H}\mathbf{w}|^{2}}{(1-\theta)\sigma_{e}^{2}} \bigg) \bigg]^{+},\nonumber
%\\ ~\forall k,\forall l, \nonumber
\end{align}
and $ \mathbf{w} = \sqrt{P}\mathbf{v} $.
\begin{remark}\label{remark Feasiblility_of_power_efficiency}
Problem \eqref{eq:Power_efficient_ori_rewritten}  may be infeasible, depending on the value of  $\bar{R}$ and on the channel vectors $h_{s,k}$ and $h_{e,l}.$ However, the feasibility can be ensured with probability of 1 for the case where the users' and eavesdroppers' channels are drawn randomly and independently from a continuous distribution with full-rank covariance matrix (e.g., a proper Gaussian non-degenerate $N_T$-variate distribution) and the number of antennas $N_T$ is not smaller than $K + L$ (e.g., in the massive multiantenna systems)\footnote{In this case, it is always possible to use zero-forcing beamforming to all eavesdroppers, such that for sufficiently large $P$ any target secrecy rate $\bar{R}$ can be achieved. The condition of sufficient transmit power $P$ can be guaranteed because of the fact that the PBs and the transmitter belong to the same service provider and they can work together on an adaptive power transfer policy (i.e., $P$ is adjusted according to the required solution $\mathbf{w}$ such that $P = \|\mathbf{w}\|^2$).}.

\end{remark}

It can be observed that even when problem \eqref{eq:Power_efficient_ori_rewritten} is feasible, it still still nonconvex. 
In order to solve this problem, we considered a two-level approach, where the inner level can be handled by employing the convex conic reformulation for a given energy transfer time allocation factor $ \theta $, whereas the outer level consists of a line search over the parameter $\theta$ in (0,1). 
%is recast as a single-variable optimization problem with respect to $ \theta $.
Now, we rewrite the problem \eqref{eq:Power_efficient_ori_rewritten} into the following two levels: 
\begin{enumerate}
\item Inner level:
\begin{align}\label{eq:Power_efficient_ori_inner}
f(\theta) = \min_{\mathbf{w}} &~ \frac{\theta(\sum_{m=1}^{M} \|\mathbf{g}_{m}\|^{2})}{1-\theta} \|\mathbf{w}\|^{2},~
s.t. ~ R_{K}(\mathbf{w}) \geq \bar{R},
\end{align}
%where $ \mathbf{w}= \sqrt{P} \mathbf{v} $, since the problem \eqref{eq:Power_efficient_ori} achieve the optimality, $ p_{m} = P $.
\item Outer level:
\begin{align}\label{eq:Power_efficient_ori_outer}
\min_{\theta} ~ f(\theta),~
s.t. ~ 0 < \theta < 1.
\end{align}
\end{enumerate}
We first solve the inner problem \eqref{eq:Power_efficient_ori_inner} for a given $ \theta $, which can be equivalently modified as 
%\begin{align}\label{eq:Power_efficient_reformulated}
%\min_{\mathbf{w},\theta} &~ \|\mathbf{w}\|^{2}  \nonumber\\
%s.t. &~ \log\left(1+\frac{\theta |\mathbf{h}_{s,k}^{H}\mathbf{w}|^{2}}{(1-\theta)\sigma_{s}^{2}}\right) - \log\left(1+\frac{\theta |\mathbf{h}_{e,l}^{H}\mathbf{w}|^{2}}{(1-\theta)\sigma_{e}^{2}}\right) \geq \frac{\bar{R}}{1-\theta}, ~\forall k, l,\nonumber\\
%&~ 0\leq \theta \leq 1.
%\end{align}
%Problem \eqref{eq:Power_efficient_reformulated} is not convex with respect to $ \mathbf{w} $ and $ \theta $, and cannot be solved directly. In this paper, we propose a two-step optimization approach to optimize $ \mathbf{w} $ and $ \theta $ separately. First step is to optimize $ \mathbf{w} $ by solving the problem \eqref{eq:Power_efficient_reformulated} for a given $ \theta $, where the problem \eqref{eq:Power_efficient_reformulated} can be expressed as 
\begin{align}\label{eq:Power_efficient_for_w}
\min_{\mathbf{w}} &~ \|\mathbf{w}\|^{2}  \nonumber\\ 
s.t. &~ \log\left(1+\frac{ |\mathbf{h}_{s,k}^{H}\mathbf{w}|^{2}}{\bar{\sigma}_{s}^{2}}\right) - \log\left(1+\frac{ |\mathbf{h}_{e,l}^{H}\mathbf{w}|^{2}}{\bar{\sigma}_{e}^{2}}\right) \geq \bar{\bar{R}}, ~~\forall \;\;(k,l)
\end{align}
where $ \bar{\sigma}_{s}^{2} = \frac{(1-\theta)\sigma_{s}^{2}}{\theta\sum_{m=1}^{M}\|\mathbf{g}_{m}\|^{2}} $, $ \bar{\sigma}_{e}^{2} = \frac{(1-\theta)\sigma_{e}^{2}}{\theta\sum_{m=1}^{M}\|\mathbf{g}_{m}\|^{2}} $, and $ \bar{\bar{R}} = \frac{\bar{R}}{1-\theta} $.
The problem \eqref{eq:Power_efficient_for_w} is not convex in terms of $ \mathbf{w} $, and still cannot be solved efficiently. In order to circumvent this non-convex issue, we can introduce a new semi-definite matrix $ \mathbf{Q}_{s} = \mathbf{w}\mathbf{w}^{H} $, and the problem \eqref{eq:Power_efficient_for_w} can be relaxed as 
\begin{align}\label{eq:SDP_relaxation_results}
\min_{\mathbf{Q}_{s} \succeq \mathbf{0}} &~\textrm{Tr}(\mathbf{Q}_{s}), \nonumber\\
s.t. &~ \frac{1}{\sigma_{s}^{2}}\textrm{Tr}(\mathbf{h}_{s,k}\mathbf{h}_{s,k}^{H}\mathbf{Q}_{s}) - \frac{2^{\bar{\bar{R}}}}{\sigma_{e}^{2}} \textrm{Tr}(\mathbf{h}_{e,l}\mathbf{h}_{e,l}^{H}\mathbf{Q}_{s}) \geq 2^{\bar{\bar{R}}} - 1, ~\forall (k,l).
\end{align}
The above problem is a convex optimization framework, which can be solved by using interior-point methods \cite{boyd_B04}. It is not always possible to expect that the
optimal solution of \eqref{eq:SDP_relaxation_results} attains the optimum of the original problem \eqref{eq:Power_efficient_for_w}. However, the SDP relaxation will be tight if the optimal solution of \eqref{eq:SDP_relaxation_results} is of rank-one. Therefore, we characterize the rank property of the SDP relaxed solution via the following proposition:
\begin{proposition}\label{proposition:SDP_Rank_Proof}
\cite[Theorem 1]{Qiang_Li_ICC_Sec_Multicasting_2011} Provided the problem \eqref{eq:SDP_relaxation_results} is feasible, the optimal solution of \eqref{eq:SDP_relaxation_results} must satisfy the following rank inequality: 
	\begin{align}\label{eq:Rank_inequality}
	\textrm{rank}(\mathbf{Q}_{s}) \leq \min (K, \sqrt{K L})
	\end{align}
\end{proposition}
%\begin{IEEEproof}
%	Please refer to \cite[Theorem 1]{Qiang_Li_ICC_Sec_Multicasting_2011}.
%	\end{IEEEproof}
With \emph{Proposition} \ref{proposition:SDP_Rank_Proof}, we are able to identify the tightness of SDP relaxed solution via the following \emph{lemma}
\begin{lemma}\label{lemma:Rank_one_lemma}
\cite[Corollary 1]{Qiang_Li_ICC_Sec_Multicasting_2011} Provided the problem \eqref{eq:SDP_relaxation_results} feasible, it is guaranteed that \eqref{eq:SDP_relaxation_results} can yield a rank-one solution which exactly solves the problem \eqref{eq:Power_efficient_for_w} when either of the following conditions is satisfied:
\begin{enumerate}
		\item $ K = 1 $ and $ L \geq 1 $. 
		\item $ 1<K \leq 3 $, and $ L = 1.$ 
\end{enumerate}
\end{lemma}
By exploiting \emph{Proposition} \ref{proposition:SDP_Rank_Proof} and \emph{Lemma} \ref{lemma:Rank_one_lemma}, if $ \textrm{rank}(\mathbf{Q}_{s}) $ satisfies the rank-one condition in \emph{Lemma} \ref{lemma:Rank_one_lemma}, we can employ the eigen-decomposition for $ \mathbf{Q}_{s} $ to obtain the optimal transmit beamformer $ \mathbf{w} $. Otherwise, we need to use a rank
reduction algorithm to tackle this problem \cite{Yongwei_Huang_Rank_Reduction_TSP_2010}.
%successive convex approximation to tackle with the secrecy rate constraint in \eqref{eq:Power_efficient_for_w}. 
%In general, this problem can be relaxed as a SDP, however, the rank-one characterization to \eqref{eq:Power_efficient_for_w} is exploited to guarantee that the optimal solution to the relaxed problem is also optimal to \eqref{eq:Power_efficient_for_w}. 
%To circumvent this non-convex issue, we propose a novel convex conic reformulation to handle \eqref{eq:Power_efficient_for_w}. 
However, when rank-one condition in \emph{Lemma} \ref{lemma:Rank_one_lemma} is satisfied, we can also consider the following \emph{theorem} to directly solve the problem \eqref{eq:Power_efficient_for_w}. 
\begin{theorem}\label{theorem power efficient schur complement}
	The problem \eqref{eq:Power_efficient_for_w} can be reformulated into the following convex optimization framework when the rank-one condition in \emph{Lemma} \ref{lemma:Rank_one_lemma} is satisfied.
	\begin{align}\label{eq:Power_efficient_for_results}
	\min_{t \geq 0,\mathbf{w}} &~~ t, \nonumber\\
	s.t. &~ \left[\begin{array}{cc}
	t \\ \mathbf{w}
	\end{array} 
	\right] \succeq_{K_{(N_{T}+1)}} \mathbf{0}, ~ \left[\begin{array}{cc}
	\frac{1}{\bar{\sigma}_{s}} \mathbf{w}^{H}\mathbf{h}_{s,k}
	\\\frac{2^{\bar{\bar{R}}}}{\bar{\sigma}_{e}}\mathbf{w}^{H}\mathbf{h}_{e,l} \\
	(2^{\bar{\bar{R}}} - 1)^{\frac{1}{2}}
	\end{array}
	\right] \succeq_{K_{3}} \mathbf{0}, ~~~\forall \;\;(k,l).
	\end{align}
%	\begin{align}\label{eq:Power_efficient_for_results}
%	\min_{\mathbf{w}} &~ t_{1} \nonumber\\
%	s.t. &~ \left[\begin{array}{cc}
%	t_{1} \\ \mathbf{w}
%	\end{array} 
%	\right] \succeq_{K} \mathbf{0},  \nonumber\\
%	&~ \left[\begin{array}{cc}
%	\frac{1}{\bar{\sigma}_{s}^{2}}\mathbf{w}^{H}\mathbf{h}_{s,k} \mathbf{I} & \left[ \begin{array}{cc}
%	\frac{2^{\bar{\bar{R}}}}{\bar{\sigma}_{e}}\mathbf{w}^{H}\mathbf{h}_{e,l} \\
%	(2^{\bar{\bar{R}}} - 1)^{\frac{1}{2}}
%	\end{array}\right] \\
%	\left[ \begin{array}{cc}
%	\frac{2^{\bar{\bar{R}}}}{\bar{\sigma}_{e}}\mathbf{w}^{H}\mathbf{h}_{e,l} &
%	(2^{\bar{\bar{R}}} - 1)^{\frac{1}{2}}
%	\end{array}\right]  & \frac{1}{\bar{\sigma}_{s}}\mathbf{w}^{H}\mathbf{h}_{s,k}
%	\end{array} \right] \succeq \mathbf{0}, ~\forall k, \forall l.
%	\end{align}
	\end{theorem} 
\begin{IEEEproof}
Please refer to Appendix \ref{appendix therorem schur complement}.
\end{IEEEproof}
By exploiting \emph{Theorem} \ref{theorem power efficient schur complement}, it is easily verified that \eqref{eq:Power_efficient_for_results} is a convex optimization problem, which can be solved by using interior-point methods \cite{boyd_B04}. Thus, the optimal transmit power of each PB can be adaptively updated as $ P^{\textrm{opt}} = \|\mathbf{w}\|^2 $, and the optimal transmit beamforming vector $ \mathbf{v}^{\textrm{opt}} $ can also be easily achieved. Now, we consider the computation complexity of solving problem \eqref{eq:Power_efficient_for_results}. According to the analysis of the basic complexity elements in \cite{Kunyu_Wang_TSP_2014_Outage}, problem \eqref{eq:Power_efficient_for_results} includes one second-order cone (SOC) constraint with dimension $ N_T + 1,$ $ KL $ SOC constraints with dimension $ N_T $, and one linear constraint. Thus, its computation complexity can be given by 
$
\mathcal{O}\bigg( \sqrt{2 KL+3} n [ KL N_{T}^2 + (N_T+1)^2 + 1 + n^2] \bigg) \ln(\frac{1}{\epsilon}),
$
where $ n = \mathcal{O}(N_T + 1) $, and $ \epsilon > 0 $ denotes the accuracy requirement.\\
\textit{A special case}: Consider the case with a single legitimate user and a single eavesdropper only. The closed-form solution can be derived by exploiting Lagrange dual problem and KKT conditions. 
%\textcolor{blue}{Unlike \cite{Ulukus_ISIT_2007_MISO_Sec}, where it aimed to maximize the secrecy rate subject to the transmit power budget, and the optimal beamforming vector was designed based on Rayleigh quotient, whereas, in our paper, a closed-form solution can be derived based on Lagrange dual problem and KKT conditions.} 
For notational convenience, we replace the channel notations $ \mathbf{h}_{s,k} $ and $ \mathbf{h}_{e,l} $ by $ \mathbf{h}_{s} $ and $ \mathbf{h}_{e} $, respectively. The following \emph{lemma} is introduced:
\begin{lemma}\label{lemma Special_case_power_efficient}
	The optimal solution to \eqref{eq:Power_efficient_ori_inner} with only single legitimate user and single eavesdropper is given by 
	\begin{align}\label{eq:Closed_form_solution_for_special_case}
&~	\mathbf{w}^{\textrm{opt}} = \sqrt{P^{\textrm{opt}}} \mathbf{v}^{\textrm{opt}}, ~\mathbf{v}^{\textrm{opt}} = \frac{\bar{\mathbf{w}}}{\|\bar{\mathbf{w}}\|_{2}},~ \bar{\mathbf{w}} = v_{\textrm{max}} (\frac{1}{\bar{\sigma}_{s}^{2}}\mathbf{h}_{s}\mathbf{h}_{s}^{H}-\frac{2^{\bar{\bar{R}}}}{\bar{\sigma}_{e}^{2}}\mathbf{h}_{e}\mathbf{h}_{e}^{H}), \nonumber\\
&~ P^{\textrm{opt}} = \alpha^{\textrm{opt}}(2^{\bar{\bar{R}}} - 1),~ \alpha^{\textrm{opt}} = \frac{1}{\varrho_{\textrm{max}} (\frac{1}{\bar{\sigma}_{s}^{2}}\mathbf{h}_{s}\mathbf{h}_{s}^{H}-\frac{2^{\bar{\bar{R}}}}{\bar{\sigma}_{e}^{2}}\mathbf{h}_{e}\mathbf{h}_{e}^{H})}.
	\end{align}
	\end{lemma}
	\begin{IEEEproof}
		See Appendix \ref{appendix Special_case_power_efficient}.
		\end{IEEEproof}
%		By exploiting \emph{Lemma} \ref{lemma Special_case_power_efficient}, the optimal power allocation can be achieved via \eqref{eq:Closed_form_solution_for_special_case}. \\
Now, it is natural that a question may arise with regard how to tackle with the problem \eqref{eq:Power_efficient_for_w} when the rank-one condition in \emph{Lemma} \ref{lemma:Rank_one_lemma} is not satisfied. Hence, we consider a SCA based scheme to reformulate the problem \eqref{eq:Power_efficient_for_w} for any general case, yielding an SOCP. We equivalently rewrite the problem \eqref{eq:Power_efficient_for_w} by introducing a new set of variables $(x_{s,k},y_{s,k},b_{s,k}), ~\forall k$ as
\begin{subequations}
\begin{align}
\min_{\mathbf{w}, b_{s,k}} &~ \|\mathbf{w}\|^{2}, \nonumber\\
s.t. &~ x_{s,k}^{2} + y_{s,k}^{2} \geq b_{s,k}, \label{eq:Sec_constraint_general1}\\ &~ (2^{\bar{\bar{R}}} - 1)\sigma_{s}^{2} + \frac{2^{\bar{\bar{R}}}\sigma_{s}^{2}}{\sigma_{e}^{2}} |\mathbf{w}^{H}\mathbf{h}_{e,l}|^{2} \leq b_{s,k}, \label{eq:Sec_constraint_general2}\\
&~ x_{s,k} = \Re\left\{ \mathbf{w}^{H}\mathbf{h}_{s,k} \right \},~ y_{s,k} = \Im \left\{  \mathbf{w}^{H}\mathbf{h}_{s,k} \right\},  \label{eq:Sec_constraint_general3}
\end{align}
\end{subequations}
where $ (x_{s,k}, y_{s,k}, b_{s,k}) \in \mathbb{R} $, $ \forall k $ In the above reformulation, it is observed that both constraint \eqref{eq:Sec_constraint_general1} and \eqref{eq:Sec_constraint_general2} are still not convex while \eqref{eq:Sec_constraint_general3} are linear constraints. In order to further process these non-convex constraints, we first introduce iterative successive approximation methods to tackle with \eqref{eq:Sec_constraint_general1}. Specifically, set $ \mathbf{q}_{s,k} = [x_{s,k}~ y_{s,k}]^{T} $, and denote the value of this vector at the $ n $-th iteration as $ \mathbf{q}_{s,k}^{(n)} $, we consider the first-order Taylor series to approximate the left hand side of \eqref{eq:Sec_constraint_general1} as 
\begin{align}\label{eq:Approximated_for_general1}
x_{s,k}^{2} + y_{s,k}^{2} = \mathbf{q}_{s,k}^{T} \mathbf{q}_{s,k} \approx \|\mathbf{u}_{s,k}^{(n)}\|^{2} + 2 \sum_{i = 1}^{2} \mathbf{u}_{s,k}^{(n)} [\mathbf{q}_{s,k}(i) - \mathbf{u}_{s,k}^{(n)} (i)],
\end{align}
where the parameter vector $ \mathbf{u}_{s,k}^{(n)} (i) $  can be updated $ \mathbf{u}_{s,k}^{(n+1)} = \mathbf{q}_{s,k}^{(n)},~\forall k $ at  the $ (n+1) $-th iteration. From this update, it is easily verified that $ \mathbf{u}_{s,k}^{(n)} $ can be determined by $ \mathbf{q}_{s,k}^{(n-1)} $.
%denotes the $i$-th element of the vector $ \mathbf{u}_{s,k} $ and $ \mathbf{u}_{s,k}^{(n)} $, respectively.
 Thus, \eqref{eq:Approximated_for_general1} can be given by 
\begin{align}\label{eq:Approximated_sec_constraint_general1}
\|\mathbf{u}_{s,k}^{(n)}\|^{2} + 2 \sum_{i = 1}^{2} \mathbf{u}_{s,k}^{(n)} [\mathbf{q}_{s,k}(i) - \mathbf{u}_{s,k}^{(n)} (i)] \geq b_{s,k}.
\end{align}
To proceed, the constraint \eqref{eq:Sec_constraint_general2} can be equivalently reformulated into the following SOC
\begin{align}\label{eq:SOC_sec_constraint_general2}
\left\| \left[\begin{array}{cc}
(2^{\bar{\bar{R}}} - 1)^{\frac{1}{2}}\sigma_{s} \\ \frac{2^{\frac{\bar{\bar{R}}}{2}}\sigma_{s}}{\sigma_{e}} \mathbf{w}^{H}\mathbf{h}_{e,l} 
\end{array}\right] \right\|^{2} \leq b_{s,k} \Rightarrow \left\| \left[ \begin{array}{ccc}
(2^{\bar{\bar{R}}} - 1)^{\frac{1}{2}}\sigma_{s} \\ \frac{2^{\frac{\bar{\bar{R}}}{2}}\sigma_{s}}{\sigma_{e}} \mathbf{w}^{H}\mathbf{h}_{e,l} \\  \frac{(b_{s,k} - 1)}{2}
\end{array} \right] \right\| \leq \frac{(b_{s,k} + 1)}{2}
\end{align} 
\begin{remark}\label{remark:SCA_remark}
	The convexity of the term  $ \mathbf{q}_{s,k}^{T}\mathbf{q}_{s,k} $ and the first-order Taylor approximation ensures that the right hand side in \eqref{eq:Approximated_for_general1} bounds the left
	side in each iterative procedure. In other
	words, the optimal solution of the problem with the approximate constraint in \eqref{eq:Approximated_sec_constraint_general1} definitely belongs to the feasible set of the original optimization
	problem at each iteration. Also, due to the above update, in the $ (n + 1) $ step, the approximation in \eqref{eq:Approximated_sec_constraint_general1} holds with
	equality. In addition, the gradients of both sides with respect
	to the optimization variables in \eqref{eq:Approximated_sec_constraint_general1} are also the same at the $ (n+1) $-th iteration, which can prove that the solution
	of the iterative procedure satisfies the KKT conditions of the original problem \eqref{eq:Power_efficient_for_w}. This fact has been proved in \cite{Barry_Marks_Approx_Noncvx_Math_Programs_OR_1978}.
\end{remark}
By exploiting \emph{Remark} \ref{remark:SCA_remark}, the problem \eqref{eq:Power_efficient_for_w} takes the following form at the $ n $-th iteration
\begin{align}\label{eq:SCA_SOCP_results}
\min_{\mathbf{w},b_{s,k}}  &~ \|\mathbf{w}\|^{2}, \nonumber\\
s.t. &~ \eqref{eq:Approximated_sec_constraint_general1}, \eqref{eq:SOC_sec_constraint_general2}, \eqref{eq:Sec_constraint_general3}, ~\forall (k,l).
\end{align}
% The following \emph{theorem} is required to reformulate \eqref{eq:Power_efficient_for_w}.
%\begin{theorem}
%\begin{align}
%\min_{\mathbf{w}, }
%\end{align}
%\end{theorem}
Based on the above discussion, an iterative algorithm to approximately solve the problem in \eqref{eq:Power_efficient_for_w} is summarized as in \textbf{Algorithm} \ref{algorithm:SCA_algorithm}.
\vspace{0.5em}
\hrule
\vspace{0.5em}
\begin{algorithm}\label{algorithm:SCA_algorithm}
	\vspace{0.5em}
Successive convex approximation to solve \eqref{eq:Power_efficient_for_w} 
\hrule
\vspace{0.5em}
\begin{enumerate}
	\item \textbf{Initialization}: Randomly generate $ \mathbf{u}_{s,k}^{(0)}, ~\forall k $ to make \eqref{eq:SCA_SOCP_results} feasible
%	, and set iteration number $ n = 0 $.
	\item \textbf{Repeat}
	\begin{enumerate}
		\item \textbf{Solve} \eqref{eq:SCA_SOCP_results}.
		\item \textbf{Set} $ \mathbf{u}_{s,k}^{(n+1)} = \mathbf{q}_{s,k}^{(n)} $, $ \forall k $. 
		\item \textbf{Set} $ n := n + 1 $.
	\end{enumerate}
\item \textbf{Until} required accuracy is achieved or the maximum
	number of iterations is reached.
\end{enumerate}
\end{algorithm}
\hrule 
\vspace{0.5em}
From \textbf{Algorithm} \ref{algorithm:SCA_algorithm}, the initialized vector $ \mathbf{u}_{s,k} $ is given by random generation to guarantee the feasibility of \eqref{eq:SCA_SOCP_results}, which can be
updated at each iteration until $ \mathbf{u}_{s,k}^{(n+1)} = \mathbf{q}_{s,k}^{(n)} $ holds when the
algorithm converges. In addition, it is guaranteed that \textbf{Algorithm} \ref{algorithm:SCA_algorithm} converges
to a locally optimal solution (quite close to the globally
optimal solution) \cite{LN_Tran_SPL_Fast_converge_2012,LN_Tran_SPL_Large_Scale_2014}. \\
 Next, we focus on the outer level problem in \eqref{eq:Power_efficient_ori_outer}. This is a single-variable optimization problem with respect to $ \theta $, which is rewritten as follows:
\begin{align}\label{eq:Second_step_ori}
\min_{\theta} &~ f(\theta), ~
s.t. ~ 
%(1-\theta) \log\left(\frac{1-\theta + \theta t_{s,k}}{1-\theta + \theta t_{e,l}}\right) \geq \bar{R},
 0 < \theta < 1.
\end{align}
%where $ t_{s,k} = \frac{|\mathbf{h}_{s,k}^{H}\mathbf{w}|^{2}}{\sigma_{s}^{2}} $ and $ t_{e,l} = \frac{|\mathbf{h}_{e,l}^{H}\mathbf{w}|^{2}}{\sigma_{e}^{2}} $. 
We can show that $ f(\theta) $ is a convex optimization problem with respect to $ \theta $ via the following \emph{lemma}:
\begin{lemma}\label{lemma Second_step_convex_show}
%$ f(\theta) $ 
\eqref{eq:Second_step_ori} is a convex problem with respect to $ \theta $.
	\end{lemma}
	\begin{IEEEproof}
See Appendix \ref{appendix Second_step_convex_show}.
\end{IEEEproof}
By exploiting \emph{Lemma} \ref{lemma Second_step_convex_show}, the optimal energy transfer time allocation factor (i.e., $ \theta^{\textrm{opt}} $) can be obtained by using one-dimensional (1D) single-variable search (e.g., golden search) in the interval $ \theta \in [0,1].$ Then, the optimal power allocation policy can be achieved.

\section{Game theory based Secure WPCN Multiantenna Multicasting System}\label{section Game_theory_based_secure_WPCN_MISO_multicasting_system}
In the previous section, we considered that the transmitter and the PBs belong to the same service provider, in which they try to work together to achieve their common target. However, this is not always the case so that, in this section, we consider the opposite scenario where the transmitter and the PBs are from two different service providers. % as shown in Fig. \ref{fig:System_model_2}.  
Both parties want to maximize their own benefit. To model this scenario, we assume that the transmitter will have to pay for the energy services from the PBs, whereas the PBs will consider this payment as incentives to provide wireless energy transfer service. Obviously, being able to decide what price to pay for the energy service, the transmitter can take a leading role in dictating the energy trading interaction. This fits very well the model of a \textit{Stackelberg} game, which motivates us to use this game theory to optimize both parties' benefit.
%\begin{figure}[!htbp]
%	\centering
%	\includegraphics[scale = 0.65]{Results_EPS/System_model_22.eps}
%	\caption{Game theory based WPCN for multiantenna secure multicasting system.}
%	\label{fig:System_model_2}
%\end{figure}
In this game model, the transmitter (leader) first pays for the harvested energy with an energy price to maximize its utility function. Then, the PBs (followers) optimize their transmit powers based on their released energy price to maximize their individual utility function. 
\subsection{Stackelberg Game Formulation}\label{section Stackelberg game formulation}
%In this subsection, we formulate this secure WPCN MISO multicasting system as a \emph{Stackelberg} game. 
Let $ \lambda $ denote the energy price that the transmitter will pay to the PBs. The total payment of the transmitter to the $M$ PBs, donoted by $Q_M$, is written as
\begin{align}
Q_M = \lambda \theta T \sum_{m=1}^{M} p_{m} \|\mathbf{g}_{m}\|^{2},
\end{align}
where $ p_{m} $ denotes the transmit power of the $ m $th PB. Without loss of generality, we can assume $ T = 1.$ We now define the utility function of the transmitter as follows:
\begin{align}
U_{M} = \mu R_{K} -  Q_{M},
\end{align}
where $ \mu > 0 $ is the weight per a unit of secrecy throughput, by which the transmitter uses to convert the achievable secrecy rate $R_{K}$ into the equivalent revenue.
%price per unit throughput for the transmitter.
Therefore, the leader game for the transmitter can be formulated as
\begin{align}\label{eq:Stackelberg_leader_level}
\max_{\lambda,\theta,\mathbf{v}}~  U_{M}, ~ s.t. ~ 0 < \theta < 1,~ \lambda \geq 0. 
\end{align}
At the same time, each PB can be modelled as a follower that wants to maximize its own revenue function, which is defined as follows:
\begin{align}\label{eq:Utility_PB_m}
U_{PB,m} = \theta (\lambda p_{m} \|\mathbf{g}_{m}\|^{2} - \mathcal{F}_{m}(p_{m})),
\end{align}
where $ \mathcal{F}_{m}(p_{m}) $ is used to model the cost of the $ m $-th PB per unit time for wirelessly charging the transmitter with the transmit power $ p_{m} $. In this paper, we consider the following quadratic model\footnote[1]{The quadratic model has been commonly used in the energy market to model the energy cost \cite{Schober_TSG2010_DSM_GT}.} for the cost function of the PBs:
\begin{align}\label{eq:Quadratic_function_power_market}
\mathcal{F}_{m}(x) = A_{m} x^{2} + B_{m} x
\end{align}
where $ A_{m} > 0 $ and $ B_{m} > 0 $ are the constants that can be different for each PB.  
Thus, the follower game of $ m $-th PB is given by
\begin{align}\label{eq:Stackelberg_follow_level}
\max_{p_{m}} ~ U_{PB,m},~ s.t. ~ p_{m} \geq 0.
\end{align}
Both \eqref{eq:Stackelberg_leader_level} and \eqref{eq:Stackelberg_follow_level} form a \emph{Stackelberg} game for this secure WPCN multiantenna multicasting system, where the transmitter (leader) announces an energy price, and then the PBs (followers) optimize the transmit power based on the released energy price to maximize their individual revenue functions. The solution of this \emph{Stackelberg} game can be obtained by investigating the \emph{Stackelberg} equilibrium, where
the transmitter and the PBs come to an agreement on the energy price, the transmit power of each PB and the time fraction of energy transfer duration.  Note that the deviation of either the transmitter or the PBs from the \emph{Stackelberg} equilibrium will introduce a loss in their revenue functions.
\subsection{Solution of The Proposed Stackelberg Game}
In order to derive the solution of this game, the well-known \emph{Stackelberg} equilibrium concept can be defined as follows:
\begin{definition}\label{definition Stackelberg_equilibrium}
	Let $ (\theta^{\textrm{opt}}, \lambda^{\textrm{opt}}) $ denote the solutions of problem \eqref{eq:Stackelberg_leader_level} while $ \{p_{m}^{\textrm{opt}}\} $ represents the solution of problem \eqref{eq:Stackelberg_follow_level} (here, the brackets $\{ \}$ indicate a vector that include all $p_m$'s with $\forall m$). Then, the triple-variable set ($ \theta^{\textrm{opt}}, \lambda^{\textrm{opt}},p_{m}^{\textrm{opt}} $) is a \emph{Stackelberg} equilibrium of the formulated game provided that the following conditions are satisfied
	\begin{align}
	&~U_{M}(\theta^{\textrm{opt}}, \lambda^{\textrm{opt}},\{p_{m}^{\textrm{opt}}\}) \geq U_{M}(\theta, \lambda,\{p_{m}^{\textrm{opt}}\}), \\
	&~U_{PB,m}(\theta^{\textrm{opt}}, \lambda^{\textrm{opt}},p_{m}^{\textrm{opt}}) \geq U_{PB,m}(\theta^{\textrm{opt}}, \lambda^{\textrm{opt}},p_{m}),~\forall m.
	\end{align}
	for $ 0<\theta <1 $, $ \lambda \geq 0 $, and $ p_{m} \geq 0 $, $ \forall m $.
\end{definition}
%By exploiting the \emph{Stackelberg} equilibrium definition in \emph{Definition} \ref{definition Stackelberg_equilibrium},
First, it can be observed that problem \eqref{eq:Stackelberg_follow_level} is convex with respect with $ p_{m} $ for given values of $ \lambda $ and $ \theta $.
%, since the utility function of the $m$-th PB is a quadratic function with respect to its transmit power $ p_{m} $ and the constraint is affine. 
Thus, the optimal solution is obtained as in the following theorem:
\begin{theorem}\label{theorem Optimal_p_m}
	For given $ \lambda $ and $ \theta $, the optimal solution to \eqref{eq:Stackelberg_follow_level} is given by 
	\begin{align}\label{eq:Optimal_p_m}
	p_{m}^{\textrm{opt}} = \begin{cases}
	\frac{\lambda\|\mathbf{g}_{m}\|^{2} -B_{m}}{2 A_{m}}, &~ \lambda > \frac{B_{m}}{\|\mathbf{g}_{m}\|^{2}} \\
	0,&~\lambda \leq \frac{B_{m}}{\|\mathbf{g}_{m}\|^{2}}.
	\end{cases} 
	\end{align}
\end{theorem}
\begin{IEEEproof}
	The proof of this \emph{theorem} is achieved by equating the first derivative of \eqref{eq:Utility_PB_m} to zero.
	\end{IEEEproof}
From Theorem \ref{theorem Optimal_p_m}, we can deduce the following remark: 
	\begin{remark}\label{remark all_PBs_involved}
		The optimal power allocation $ p_{m}^{\textrm{opt}} $ can only be obtained under the condition that the energy price $ \lambda $ is greater than threshold $ \frac{B_{m}}{\|\mathbf{g}_{m}\|^{2}} $.
		Thus, we divide the PBs into two sets, namely, the active and non-active PBs. The PBs who can transfer the power to the transmitter and help determine the achievable secrecy rate by using the harvested energy are called active PBs. The remaining ones are non-active PBs. Generally, according to \eqref{eq:Optimal_p_m}, we can determine these active PBs and re-index them for the total power computation. In our paper, for convenience, we assume that all PBs are active and considered to be involved in the WET phase of $ \theta T.$ 
	\end{remark}
	
By exploiting \emph{Theorem} \ref{theorem Optimal_p_m} and \emph{Remark} \ref{remark all_PBs_involved}, we replace $ p_{m} $ in \eqref{eq:Stackelberg_leader_level} with \eqref{eq:Optimal_p_m}, problem \eqref{eq:Stackelberg_leader_level} now becomes
	\begin{align}\label{eq:U_L_with_optimal_p_m}
	\max_{\lambda,\theta}&~   \mu (1-\theta) \left[ \log\bigg[ 1+ \bigg( \lambda\sum_{m=1}^{M} \frac{\|\mathbf{g}_{m}\|^{4}}{2 A_{m}} - \sum_{m=1}^{M} \frac{B_{m}\|\mathbf{g}_{m}\|^{2}}{2 A_{m}} \bigg) t_{s} \bigg] \right. \nonumber \\ 
	&~~~~~~~~~~~~ \left. - \log\bigg[ 1+ \bigg( \lambda\sum_{m=1}^{M} \frac{\|\mathbf{g}_{m}\|^{4}}{2 A_{m}} - \sum_{m=1}^{M} \frac{B_{m}\|\mathbf{g}_{m}\|^{2}}{2 A_{m}} \bigg) t_{e} \bigg] \right] \nonumber\\ 
	&~~~~~~~~~~~~ - \theta \lambda^{2} \sum_{m=1}^{M} \frac{\|\mathbf{g}_{m}\|^{4}}{2A_{m}} + \theta \lambda \sum_{m=1}^{M}\frac{B_{m}\|\mathbf{g}_{m}\|^{2}}{2A_{m}} \nonumber\\ s.t. &~ 0 < \theta < 1,~ \lambda \geq 0
	\end{align}
	where 
		\begin{align}
		t_{s} = \min_{k} t_{s,k} = \min_{k} \frac{\theta |\mathbf{h}_{s,k}^{H}\mathbf{v}|^{2}}{(1-\theta)\sigma_{s}^{2}},~\forall k, ~t_{e} = \max_{l} t_{e,l} = \min_{l} \frac{\theta |\mathbf{h}_{e,l}^{H}\mathbf{v}|^{2}}{(1-\theta)\sigma_{e}^{2}},~\forall l. \nonumber
		\end{align}	
%	\begin{align}
%	R_{k} = (1-\theta) \left[ \log \left( 1+ \frac{\theta \sum_{m=1}^{m}p_{m}^{\textrm{opt}}\|\mathbf{g}_{m}\|^{2} |\mathbf{h}_{s,k}^{H}\mathbf{v}|^{2}}{(1-\theta)\sigma_{s}^{2}} \right) - \max_{l}  ~\log \left( 1+ \frac{\theta \sum_{m=1}^{m}p_{m}^{\textrm{opt}}\|\mathbf{g}_{m}\|^{2} |\mathbf{h}_{e,l}^{H}\mathbf{v}|^{2}}{(1-\theta)\sigma_{e}^{2}} \right) \right]
%	\end{align}
Note that before solving problem \eqref{eq:U_L_with_optimal_p_m}, the normalized transmit beamfoming vector $ \mathbf{v} $ can be obtained by employing a similar approach as in Section \ref{section Power Efficient for WPCN MISO Secure System} Thus, we solve for the optimal solution to the energy price $ \lambda $ and the energy transfer time allocation $ \theta $ only in leader level game \eqref{eq:U_L_with_optimal_p_m}. 
	The problem \eqref{eq:U_L_with_optimal_p_m} is not jointly convex in terms of $ \theta $ and $ \lambda.$ It is extremely hard to find their optimal solutions simultaneously due to the complexity of the objective function in  \eqref{eq:U_L_with_optimal_p_m}. In order to address this issue, we first derive the closed-form solution for the optimal $ \lambda $ with a given value $ \theta $. Then, the optimal value for $ \theta $ can be achieved through numerical analysis. 
%	For given $ \theta $ and $ \lambda $, we consider the following problem to obtain the optimal $ \mathbf{v} $. 
%	\begin{align}
%	\max_{\mathbf{v}} ~ \min_{k} R_{k},~ s.t. ~ \mathbf{v}^{H}\mathbf{v} = 1.
%	%\|\mathbf{v}\|^{2} \leq  \lambda \theta \sum_{m=1}^{M} p_{m}^{\textrm{opt}} \|\mathbf{g}_{m}\|^{2}. 
%	\end{align}
%	The above problem can be easily relaxed as a SDP, which can be solved by interior-point method ... ...
%	\begin{align}
%    \Phi_{m} =\left\{
%     \begin{aligned}
%     1, &~  \lambda > \frac{B_{m}}{\|\mathbf{g}_{m}\|^{2}} \\
%     0, &~ \lambda \leq \frac{B_{m}}{\|\mathbf{g}_{m}\|^{2}}.
%     \end{aligned}
%     \right.
%	\end{align}
%	Special case with $ \Phi_{m} = 1 $, which indicates that all PBs send the energy to the transmitter. 
In order to derive the closed-form solution of $ \lambda $, we set $ C_{M} = \sum_{m=1}^{M} \frac{\|\mathbf{g}_{m}\|^{4}}{2 A_{m}} $, $ D_{M} = \sum_{m=1}^{M} \frac{B_{m}\|\mathbf{g}_{m}\|^{2}}{4A_{m}} $, and the objective function to \eqref{eq:U_L_with_optimal_p_m} can be expressed as 
\begin{align}\label{eq:U_L_lambda}
 U_{M}(\theta, \lambda) & = \mu (1-\theta) \left[ \log\bigg[ 1+ \bigg( \lambda C_{M} - 2 D_{M} \bigg) t_{s} \bigg] - \log\bigg[ 1+ \bigg( \lambda C_{M} - 2 D_{M} \bigg) t_{e} \bigg] \right] \nonumber\\ 
 &~~~~~~~- \theta \lambda^{2} C_{M} + 2 \theta \lambda D_{M}.
\end{align}
\begin{lemma}\label{lemma lambda_concave_function}
	\eqref{eq:U_L_lambda} is a concave function with respect to $ \lambda $.
\end{lemma}
\begin{IEEEproof}
	See Appendix \ref{appendix lambda_concave_function}.
\end{IEEEproof}	
By exploiting \emph{Lemma} \ref{lemma lambda_concave_function}, we can claim that \eqref{eq:U_L_with_optimal_p_m} is a convex problem. Now, we derive the closed-form solution of $ \lambda $.
%\begin{lemma}
%	The optimal solution of $ \lambda $ can be derived in terms of closed-form solution as follows:
%		\begin{align}
%		x^{\textrm{opt}} =  e^{j \angle x_{1}} \sqrt[3]{|x_{1}|} +  e^{j \angle x_{2}} \sqrt[3]{|x_{2}|} - \frac{a}{3}
%		\end{align}
%		where $ \angle $ denotes the phase angle of an complex random
%		variable, and 
%		\begin{align}
%		x_{1} & = -\frac{q}{2} + \sqrt{\Delta},~x_{2} = -\frac{q}{2} - \sqrt{\Delta}, \nonumber\\
%		\Delta & = \frac{p^{3}}{27} + \frac{q^{2}}{4},~
%		p = -\frac{a^{2}}{3} + b,~ q = \frac{2 a^{3}}{27} - \frac{ab}{3}+c,\nonumber\\
%		a& = \frac{D_{M}}{2} + \bigg( \frac{1}{t_{e}} + \frac{1}{t_{s}} \bigg),~ b = \frac{1}{t_{e}t_{s}} + \frac{D_{M}}{2} \bigg( \frac{1}{t_{e}} + \frac{1}{t_{s}} \bigg),~ c = \frac{D_{M}}{2t_{e}t_{s}} - \frac{\mu (1-\theta)C_{M}}{2} \bigg( \frac{1}{t_{e}} - \frac{1}{t_{s}} \bigg).
%		\end{align}
%\end{lemma} 
%\begin{IEEEproof}
In order to obtain the optimal solution to $ \lambda $, let the first order derivative of \eqref{eq:U_L_lambda} equate to zero, we have
\begin{align}
\frac{\mu (1-\theta) t_{s} C_{M} }{1+(\lambda C_{M} -  D_{M}) t_{s} - D_{M} t_{s}} - \frac{\mu (1-\theta) t_{e} C_{M} }{1+(\lambda C_{M} -   D_{M}) t_{e} - D_{M} t_{e}} - 2 \theta C_{M} \lambda + 2 \theta D_{M} = 0.
\end{align}
Set $ x = \lambda C_{M} - D_{M} $, and after a few of mathematical simplifications, we arrive at 
\begin{align}\label{eq:Cubic_equation_with_x}
%x^{3} + \bigg[ \frac{D_{M}}{2} + \bigg(\frac{1}{t_{e}} + \frac{1}{t_{s}}\bigg) \bigg] x^{2} + \bigg[ \frac{1}{t_{e}t_{s}} + \frac{D_{M}}{2} \bigg(\frac{1}{t_{e}} + \frac{1}{t_{s}}\bigg) \bigg] x + \bigg[ \frac{D_{M}}{2t_{e}t_{s}} - \frac{\mu (1-\theta) C_{M}}{2} \bigg(\frac{1}{t_{e}} - \frac{1}{t_{s}}\bigg) \bigg] = 0,
x^{3} + a x^{2} + b x + c  = 0,
\end{align}
where 
%\begin{align}
%a = \frac{2 (t_{s} + t_{e}) - 5 D_{M}t_{s}t_{e}}{2 t_{s} t_{e}}, ~ b = \frac{2 - 3 D_{M} (t_{s} + t_{e}) + 4 D_{M}^{2} t_{s} t_{e}}{2 C_{M}t_{s} t_{e}},\nonumber\\
%c = \frac{\theta D_{M}^{2}(t_{s} + t_{e}) - \theta D_{M} - \theta D_{M}^{2} t_{s}t_{e} - \mu(1-\theta)C_{M}(t_{s} - t_{e})}{2\theta C_{M}^{2} t_{s}t_{e}}.
%\end{align}
\begin{align}
a = \frac{(t_{s}+t_{e}) - 2 D_{M} t_{s}t_{e}}{t_{s} t_{e}},~ b = \frac{(D_{M} t_{s}  - 1) (D_{M} t_{e} - 1)}{t_{s} t_{e}},~ c = - \mu (1-\theta) C_{M} (t_{s} - t_{e}).\nonumber
\end{align}

 It is easily observed that \eqref{eq:Cubic_equation_with_x} is a cubic equation, which can be solved in terms of closed-form solution of $ x $ by using Cardano's formula \cite{Xiaobin_Huang_CL_2014_FD},
\begin{align}\label{eq:Optimal_lambda}
x^{\textrm{opt}} =  e^{j \angle x_{1}} \sqrt[3]{|x_{1}|} +  e^{j \angle x_{2}} \sqrt[3]{|x_{2}|} - a/3,
\end{align}
where $ \angle $ denotes the phase angle of an complex random
variable, and 
\begin{align}
x_{1} & = -\frac{q}{2} + \sqrt{\Delta},~x_{2} = -\frac{q}{2} - \sqrt{\Delta}, \nonumber\\
\Delta & = \frac{p^{3}}{27} + \frac{q^{2}}{4},~
p = -\frac{a^{2}}{3} + b,~ q = \frac{2 a^{3}}{27} - \frac{ab}{3}+c.\nonumber
%\\
%a& = \frac{D_{M}}{2} + \bigg( \frac{1}{t_{e}} + \frac{1}{t_{s}} \bigg),~ b = \frac{1}{t_{e}t_{s}} + \frac{D_{M}}{2} \bigg( \frac{1}{t_{e}} + \frac{1}{t_{s}} \bigg),~ c = \frac{D_{M}}{2t_{e}t_{s}} - \frac{\mu (1-\theta)C_{M}}{2} \bigg( \frac{1}{t_{e}} - \frac{1}{t_{s}} \bigg).
\end{align}	
Thus, we obtain the optimal energy price as 
\begin{align}\label{eq:Optimal_energy_transfer_price}
\lambda^{\textrm{opt}} = \frac{x^{\textrm{opt}} + D_{M}}{C_{M}}.
\end{align}
%Thus, the optimal energy price can be achieved as 
%\begin{align}\label{eq:Optimal_lambda}
%\lambda^{\textrm{opt}} = \frac{x^{\textrm{opt}} + D_{M}}{C_{M}}.
%\end{align}
%	\end{IEEEproof}
We have already obtained the optimal energy price of the transmitter for a given $ \theta $. Now, the optimal energy transfer time allocation is derived in the following. We substitute the closed-form expression \eqref{eq:Optimal_energy_transfer_price} into \eqref{eq:U_L_lambda}, thus, the following optimization problem can be written with respect to $ \theta $, 
\begin{align}\label{eq:U_L_with_theta}
\max_{\theta}~ U_{M}(\theta,\lambda^{\textrm{opt}}), s.t.~ 0<\theta <1.
\end{align}
The problem \eqref{eq:U_L_with_theta} can be efficiently handled by using 1D search to obtain the optimal energy transfer time solution $ \theta^{\textrm{opt}} $ as follows:
\begin{align}\label{eq:Optimal_theta}
\theta^{\textrm{opt}} = \arg\max_{\theta \in (0,1)} U_{M}(\theta,\lambda^{\textrm{opt}}). 
\end{align}
We have completed the derivation of the \emph{Stackelberg} equilibrium $(p_{m}^{\textrm{opt}},\lambda^{\textrm{opt}},\theta^{\textrm{opt}})$ for the formulated \emph{Stackelberg} game, which are shown in \eqref{eq:Optimal_p_m}, \eqref{eq:Optimal_lambda} and \eqref{eq:Optimal_theta}.

\section{Numerical Results}\label{section numerical_results}
In this section, we provide the simulation results to validate the proposed schemes. We consider the secure multiantenna multicasting system that consists of three legitimate receivers (i.e., $ K = 3 $) and two eavesdroppers ($ L = 5 $), where the transmitter is wirelessly powered by five PBs ($ M = 5 $). It is assumed that the transmitter is equipped with eight transmit antennas (i.e., $ N_{T} = 8 $), whereas the others consist of single antenna. We employ the path loss channel model $ \sqrt{A d_{x}^{-\alpha}} $, where $ A = 10^{-3}.$ The path loss exponent is set to $ \alpha = 3.$ Distance variable $ d_{x} $ can be replaced with $ d_{s} $, $ d_{e} $, and $ d_{PB} $ according to different channel coefficients, representing the distance between the transmitter and the legitimate users, the eavesdroppers as well as the PBs, respectively. In our simulation, we choose $ d_{s} = d_{e} = 2~\textrm{m} $, and $ d_{PB} = 5~\textrm{m} $ unless specified. The target secrecy rate is set to be $ \bar{R} = 2~\textrm{bps/Hz} $. The noise powers at the legitimate users and the eavesdroppers are set as $ \sigma_{s}^{2} = \sigma_{e}^{2} = 10^{-8}~ \textrm{mW} $.  

\begin{figure}[!htbp]
	\centering
	\includegraphics[scale = 0.85]{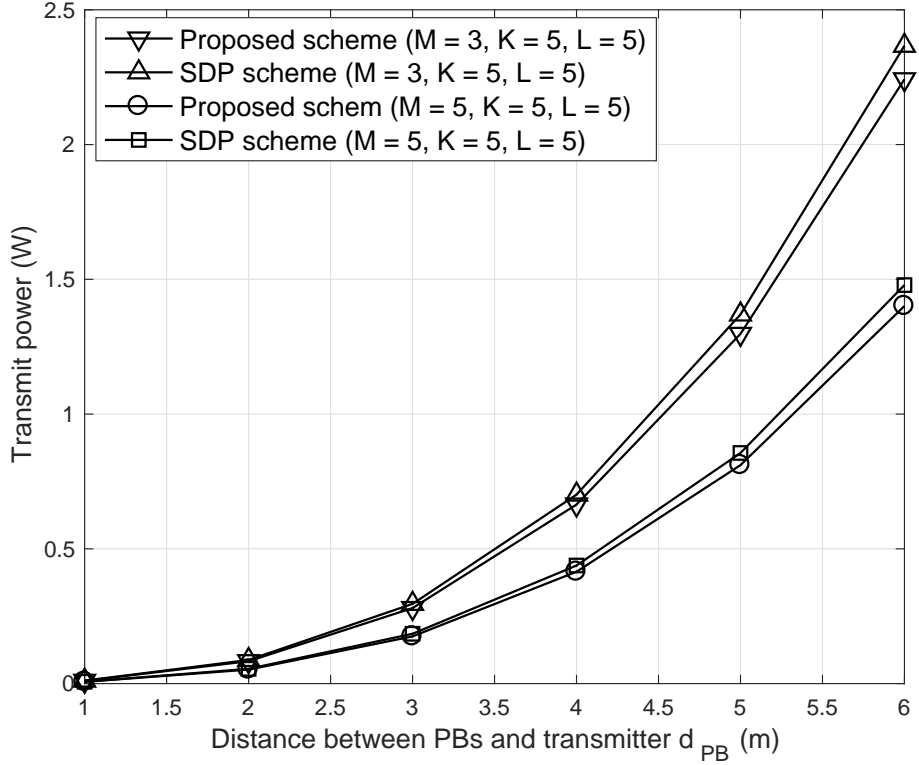}
	\caption{Comparison of transmit power between the proposed scheme and SDP scheme versus distance between PBs and transmitter.}
	\label{fig:Power_vs_distance_K5L5_general_case}
\end{figure}

\begin{figure}[!htbp]
	\centering
	\includegraphics[scale = 0.85]{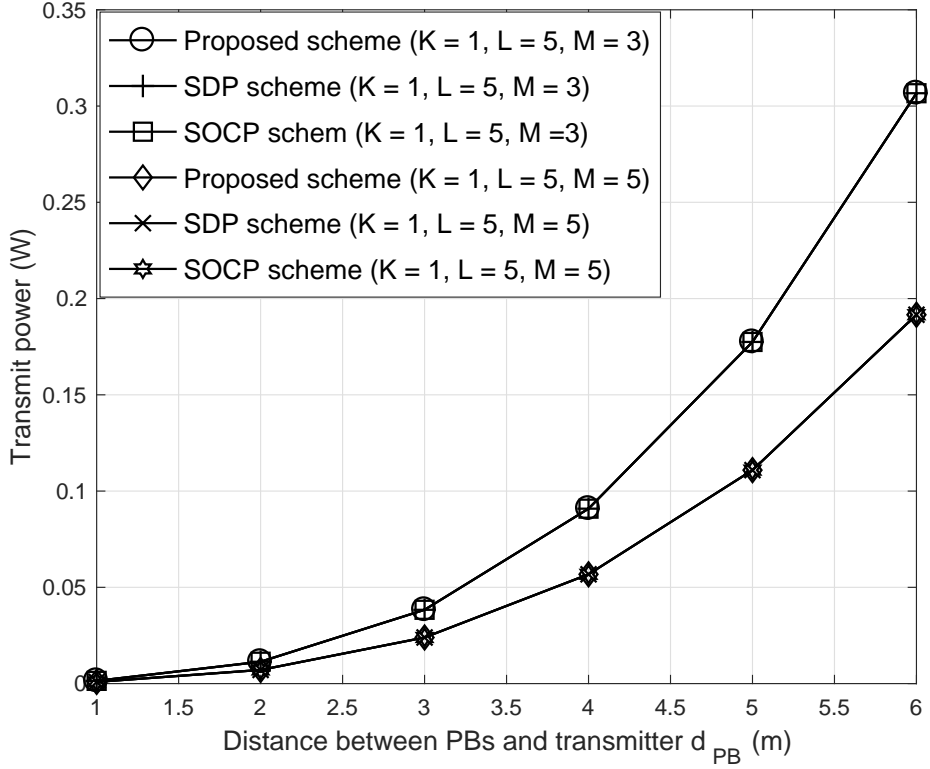}
	\caption{Comparison of transmit power between the SOCP and SDP schemes versus distance between PBs and transmitter.}
	\label{fig:Power_vs_distance_K1L5_satisfying_rank_condition}
\end{figure}

\begin{figure}[!htbp]
	\centering
	\includegraphics[scale = 0.7]{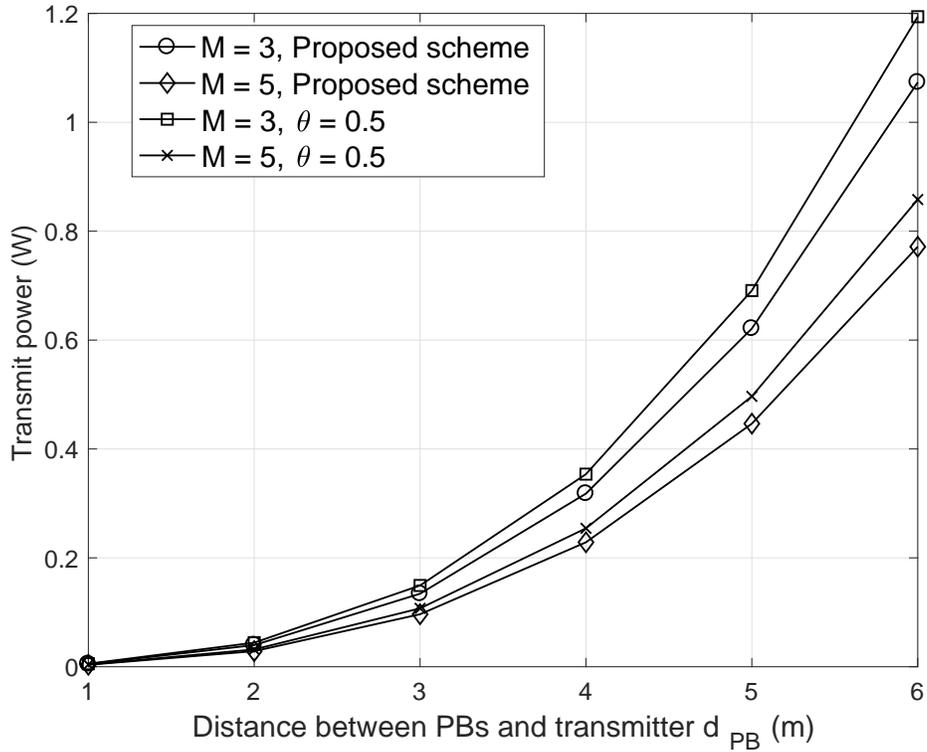}
	\caption{Comparison of transmit power between the proposed scheme with optimal $ \theta $ and fixed $ \theta $ versus distance between PBs and transmitter.}
	\label{fig:Power_vs_distance_special_02}
\end{figure}
\begin{figure}[!htbp]
	\centering
	\includegraphics[scale = 0.7]{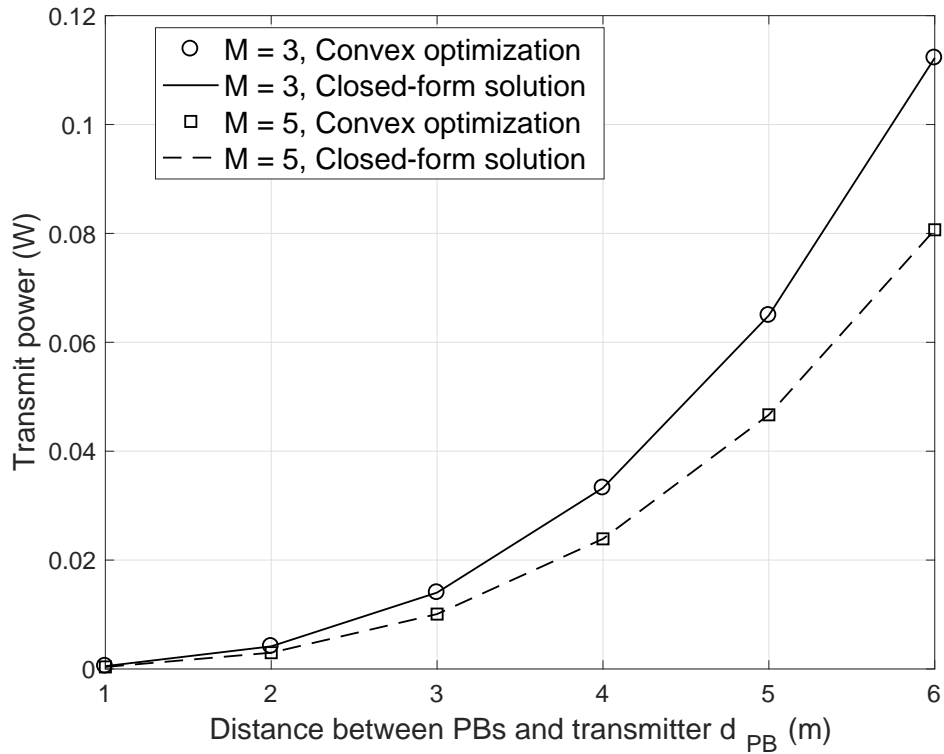}
	\caption{Transmit power versus distance between PBs and transmitter with special case.}
	\label{fig:Power_vs_distance_special_01}
\end{figure}

\begin{figure}[!htbp]
	\centering
	\includegraphics[scale = 0.8]{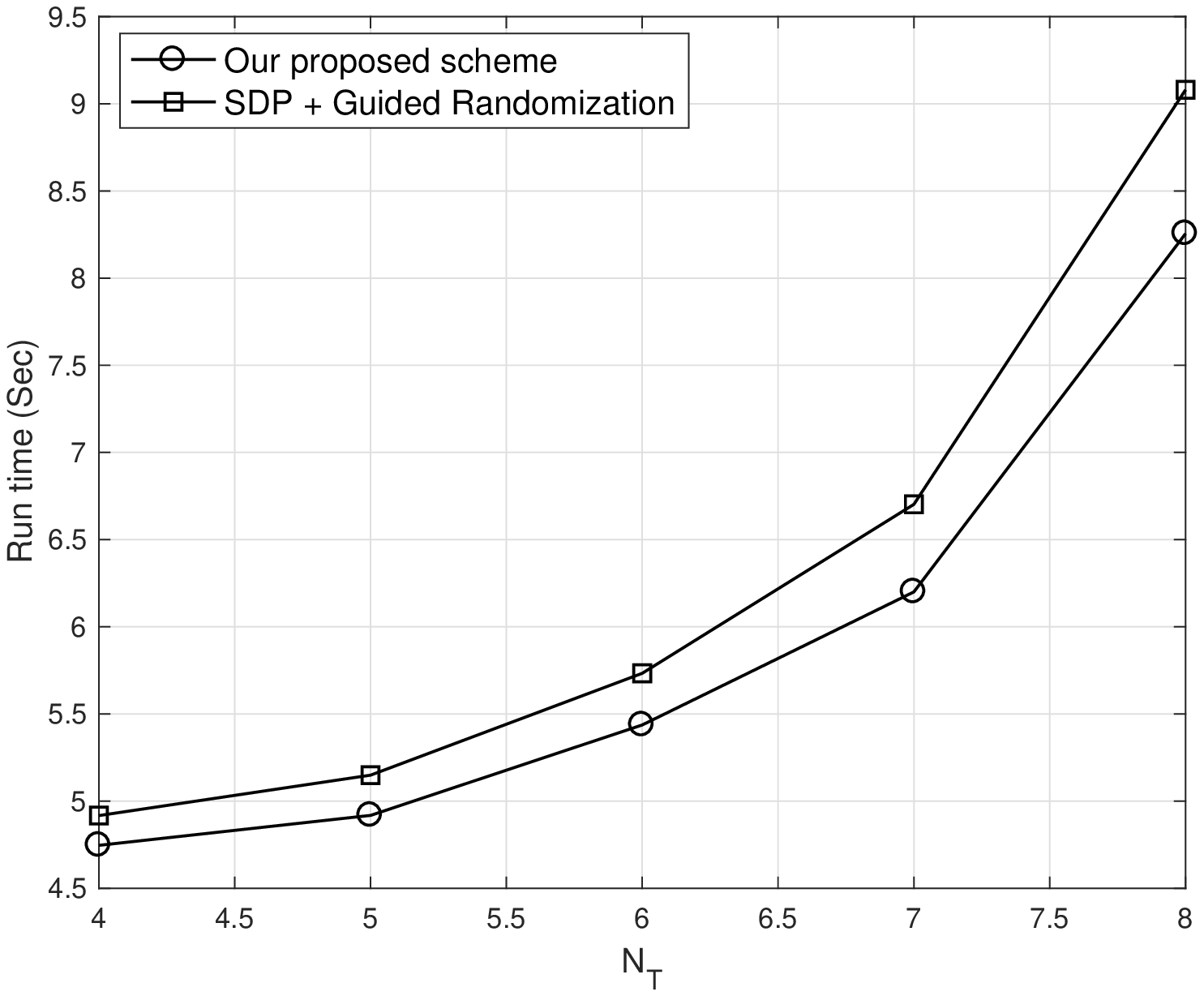}
	\caption{Run time versus the number of transmit antenna $ N_{T} $.}
	\label{fig:Run_time_comparison}
\end{figure}
First, we evaluate the minimized transmit power obtained from our proposed scheme in Section \ref{section Power Efficient for WPCN MISO Secure System} against the distance between PBs and the transmitter (i.e., $ d_{PB} $).  Fig. \ref{fig:Power_vs_distance_K5L5_general_case} shows the result with general case when the rank-one condition in \emph{Lemma} \ref{lemma:Rank_one_lemma} is not satisfied. From this result, one can observe that our proposed SCA scheme outperforms the SDP scheme in \cite{Cumanan_JSTSP_2016_Game}. This is owing to a fact that our proposed scheme can achieve a optimal solution via \textbf{Algorithm} \ref{algorithm:SCA_algorithm}, whereas the SDP scheme cannot satisfy the rank-one condition such that the relaxed solution cannot achieve optimality. Fig. \ref{fig:Power_vs_distance_K1L5_satisfying_rank_condition} evaluates the results of \emph{Theorem} \ref{theorem power efficient schur complement}, where it is observed that the SOCP yields the same performance with the proposed SCA scheme and SDP scheme, which confirms that the correctness and accuracy of \emph{Theorem} \ref{theorem power efficient schur complement}.
	Fig. \ref{fig:Power_vs_distance_special_02} shows the impact of the energy time allocation.
%	the results for the case ($ K = 2 $ and $ L = 1 $) with either three ($M=3$) or five ($M = 5$) PBs. 
%	It is observed that the minimized transmit power increases with the distance $ d_{PB},$ which means that the PBs should be placed near to the transmitter such that less transmit power is required to guarantee secure transmission.  
%	%obviously owing to the fact that the nearer the PBs to the transmitter, the higher the energy transfer efficiency from the PBs to the transmitter which reduces the transmit power. 
%It is also observed that the larger number of PBs involved would lead to lower amount of transmit power required. 
From this figure, we observe that the proposed SOCP scheme with the fixed time allocation based scheme (i.e., $ \theta = 0.5 $) that obviously requires more transmit power than the proposed SOCP scheme with optimal energy time allocation, this is owing to a fact that our proposed scheme can achieve a optimal energy time allocation by numerical search (i.e., $ \theta = \theta^{\textrm{opt}} $).  
Similar behaviours are observed in Fig. \ref{fig:Power_vs_distance_special_01} which is obtained for the special case of single user and single eavesdropper. This figure shows that the derived closed-form solution in \emph{Lemma} \ref{lemma Special_case_power_efficient} matches well with the numerical results obtained from a convex optimization tool, which validates the accuracy of this closed-form solution. 
In Fig. \ref{fig:Run_time_comparison}, we compare the run time of our proposed scheme and the SDP relaxation with \emph{Guided Randomzation}\footnote{\emph{Guided Randomization} is employed to tackle with the scenario that the SDP relaxed solution is not rank-one.} \cite{Jialing_Liao_CL_Robust_PS_SWIPT_2016} versus the number of transmit antenna $ N_{T} $. From this result, one can observe that our proposed scheme consume less time than SDP with \emph{Guided Randomization}, which implies that proposed algorithm has a lower computational complexity.  

%Next, we discuss the simulation results in Fig. \ref{fig:Power_vs_distance_special_01} to support the closed-form solution derived in \emph{Lemma} \ref{lemma Special_case_power_efficient} for the special case with $ K = 1 $ and $ L = 1 $. As mentioned in Section \ref{section Power Efficient for WPCN MISO Secure System}, the power minimization problem can be solved by convex optimization framework and closed-form solution. Fig. \ref{fig:Power_vs_distance_special_01} shows that the transmit power performance for this special case, where it is observed from this figure that both convex optimization framework and closed-form solution are the same such that shows the accuracy of the closed-form solution. Then, Fig. \ref{fig:Power_vs_distance_special_02} shows the transmit power performance of the general case (i.e., $ K = 3, L = 2 $). We compare our proposed scheme with the fixed time allocation based scheme (i.e., $ \theta = 0.5 $), in which it is easily shown that our proposed scheme has a less power consumption than the fixed time allocation based scheme. }
\begin{figure}[!htbp]
	\centering
	\includegraphics[scale = 0.7]{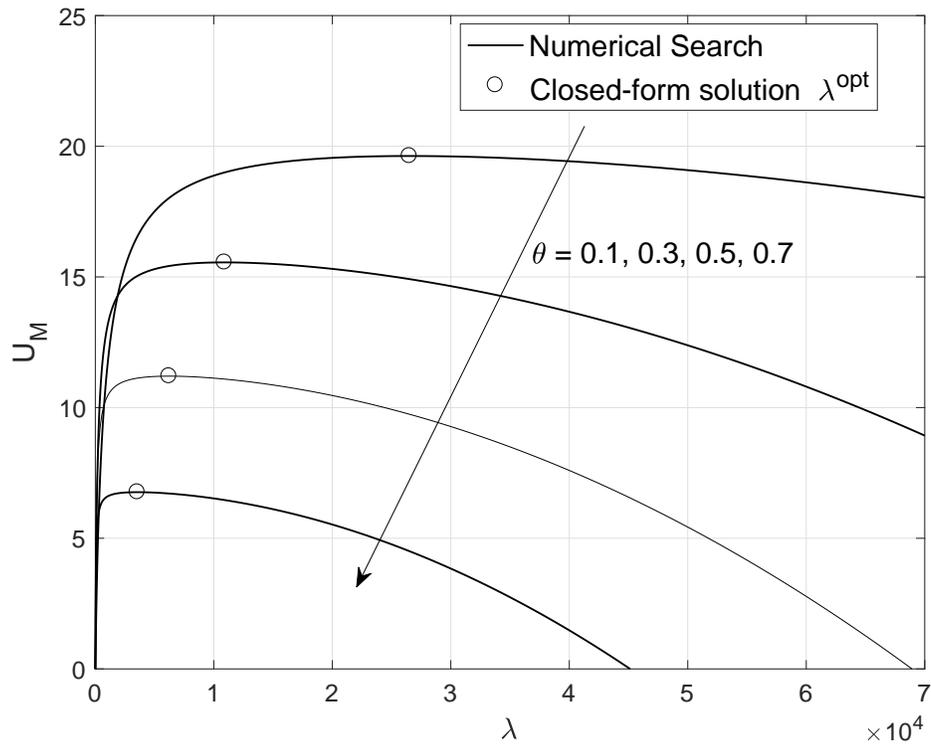}
	\caption{The utility function of transmitter $ U_{M} $ versus energy transfer price $\lambda$}
	\label{fig:U_L_vs_lambda}
\end{figure}
\begin{figure}[!htbp]
	\centering
	\includegraphics[scale = 0.7]{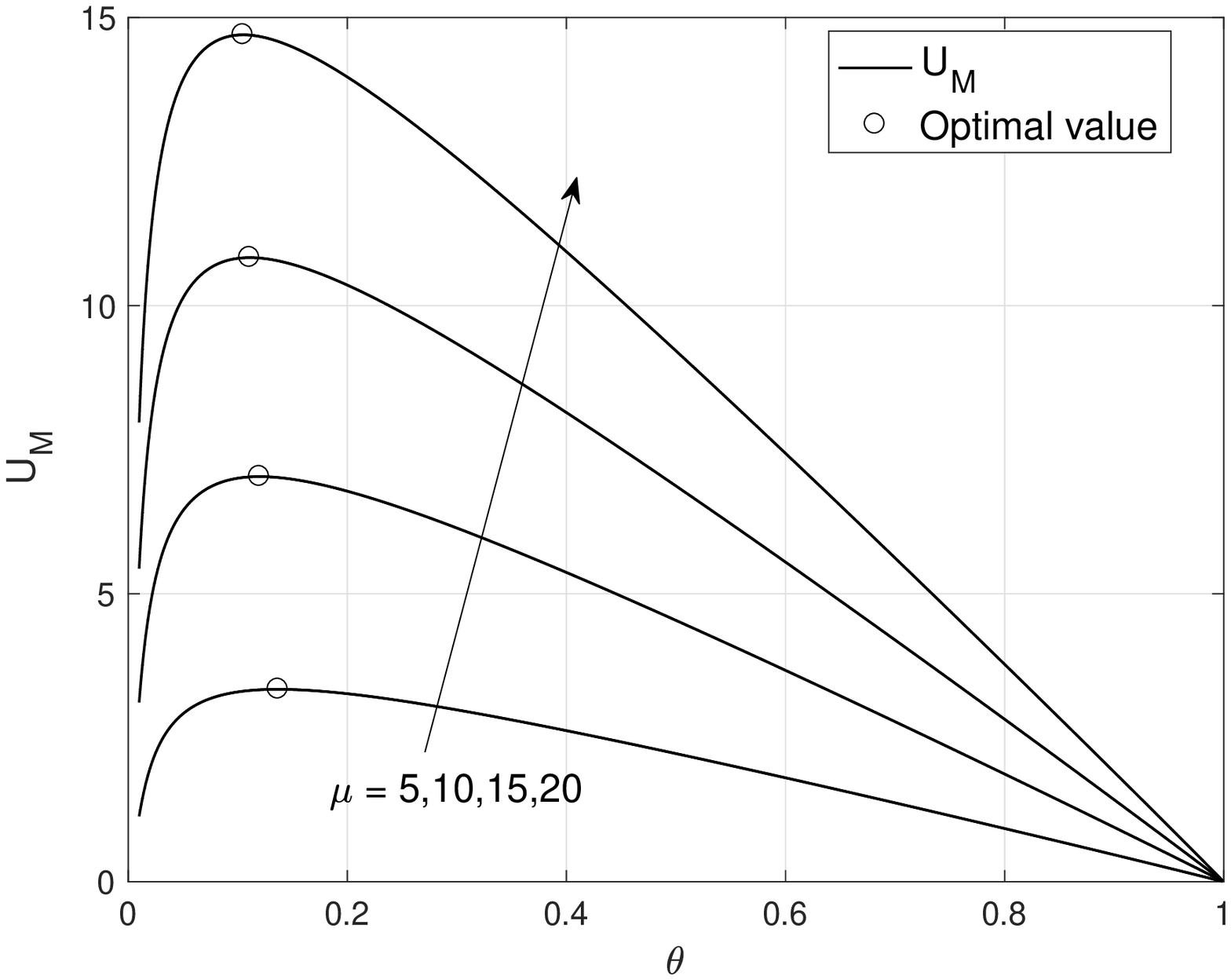}
	\caption{The utility function of transmitter $ U_{M} $ versus energy transfer time allocation $\theta$}
	\label{fig:U_L_vs_theta}
\end{figure}

Next, we validate the equilibrium of the proposed
\emph{Stackelberg} game. In order to support the derived \emph{Stackelberg} equilibrium, we first evaluate the utility function of the transmitter versus the energy transfer price $ \lambda $ with a fixed energy transfer time allocation $ \theta $ in Fig. \ref{fig:U_L_vs_lambda}. From this figure, it is observed that the revenue function is concave, which validates the proof of convexity shown in \emph{Lemma} \ref{lemma lambda_concave_function}. In this figure, it also can be shown that the optimal utility function of the transmitter can be obtained via optimal energy transfer price $ \lambda^{\textrm{opt}} $ in \eqref{eq:Optimal_energy_transfer_price} and it matches the numerical search with different given $ \theta $, which confirms the optimal closed-form solution of the energy transfer price $ \lambda $. Also, as $ \theta $ increases, the utility function of the legitimate transmitter is decreasing, and the optimal value $ \lambda $ shifts to the left.% \textcolor{blue}{which means that the less energy price paid can achieve the optimal revenue of the transmitter. }
%	This is owing to a fact that the more energy time would lead to larger energy transfer such that faster optimal utility function of legitimate transmitter is achieved with less energy price.
In addition, the revenue function of the transmitter versus energy transfer time allocation (i.e., $ \theta $) with optimal energy price $ \lambda^{\textrm{opt}} $ is shown in Fig. \ref{fig:U_L_vs_theta}. From this figure, it is shown that the revenue function is concave with respect to $ \theta $, which validates \eqref{eq:Optimal_theta}. Moreover, there exists a optimal utility transfer time (i.e., $  \theta^{\textrm{opt}} $) via numerical search with the optimal energy price. As $ \mu $ increases, the optimal value slightly shifts to the left. % \textcolor{blue}{which means that the less energy time consumption can achieve the optimal revenue of the transmitter. }

\begin{figure}[!htbp]
	\centering
	\includegraphics[scale = 0.7]{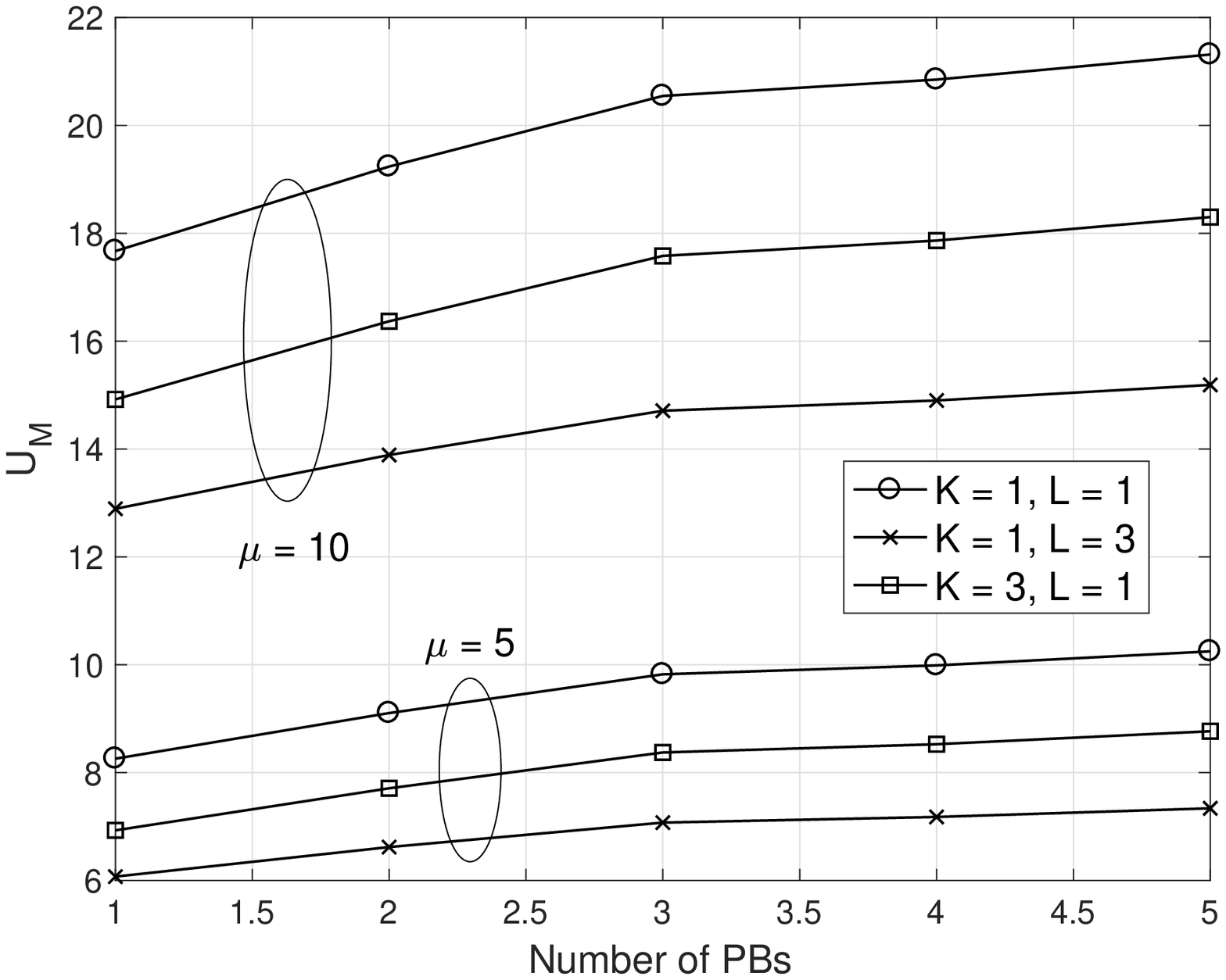}
	\caption{The utility function of transmitter $ U_{M} $ versus No. of PBs with different number of legitimate users and eavesdroppers}
	\label{fig:U_L_vs_No_PB}
\end{figure}

\begin{figure}[!htbp]
	\centering
	\includegraphics[scale = 0.7]{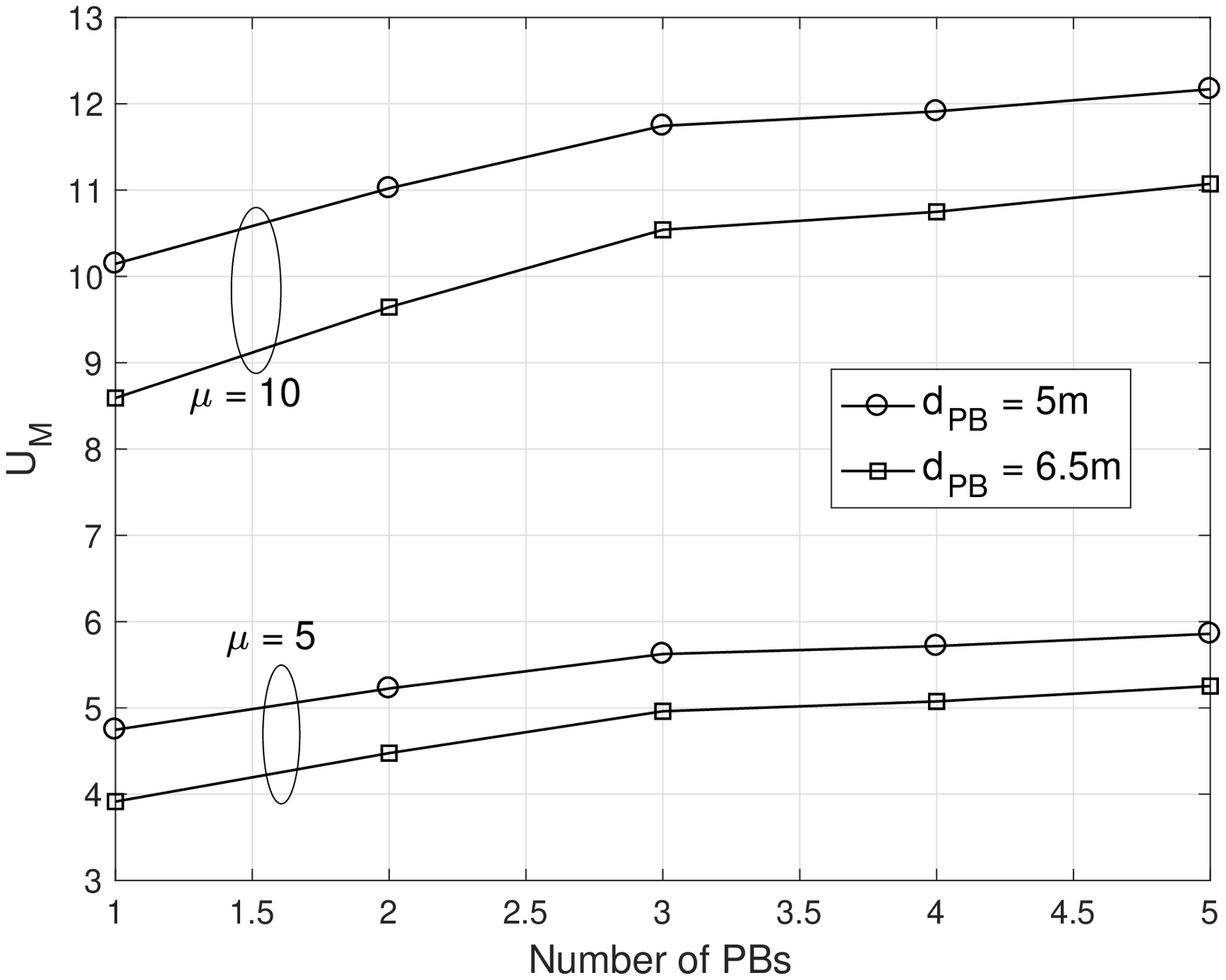}
	\caption{The utility function of transmitter $ U_{M} $ versus No. of PBs with different $ d_{PB} $}
	\label{fig:U_L_vs_No_PB_1}
\end{figure}

\begin{figure}[!htbp]
	\centering
	\includegraphics[scale = 0.7]{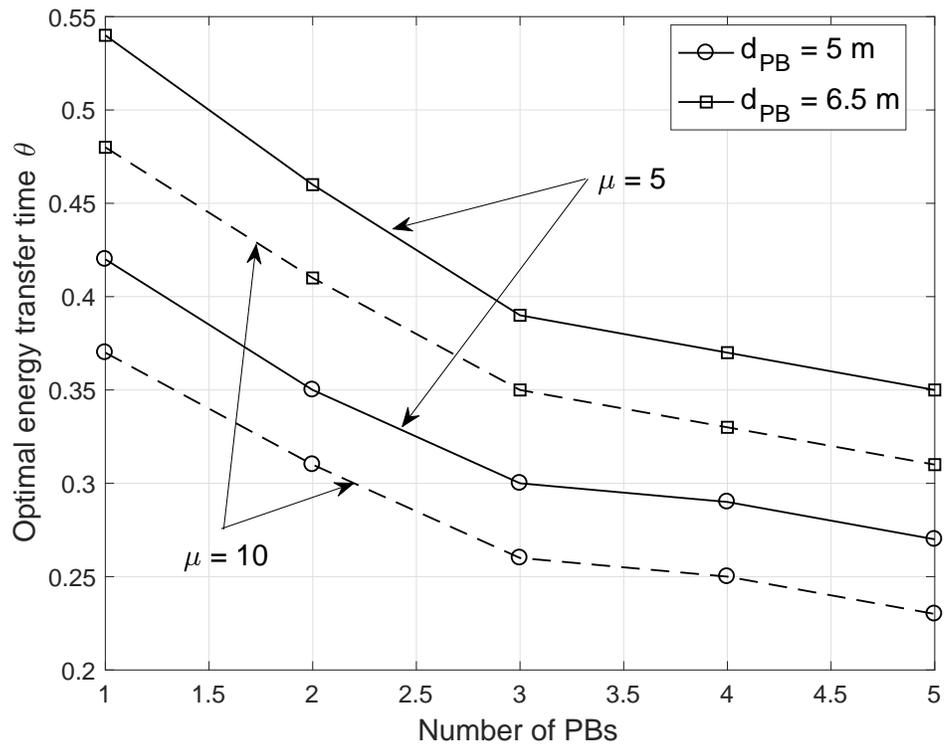}
	\caption{Optimal energy transfer time allocation $ \theta $ versus No. of PBs}
	\label{fig:Theta_vs_No_PB}
\end{figure}

\begin{figure}[!htbp]
	\centering
	\includegraphics[scale = 0.7]{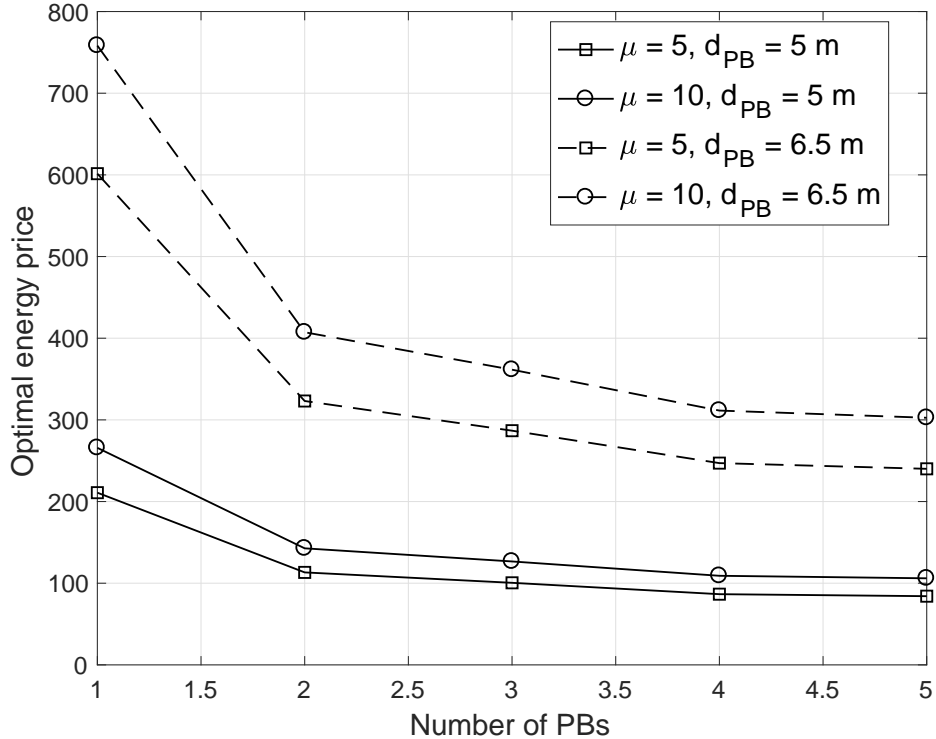}
	\caption{The optimal energy price versus No. of PBs.}
	\label{fig:Optimal_price_vs_No_PB}
\end{figure}

Then, we evaluate the transmitter revenue function performance of the proposed \emph{Stackelberg} game. 
Fig. \ref{fig:U_L_vs_No_PB} and Fig. \ref{fig:U_L_vs_No_PB_1} show the revenue function of the transmitter versus the number of PBs. From both figures, we can observe that this utility is improved with increasing of the number of PBs and $ \mu $. 
From Fig. \ref{fig:U_L_vs_No_PB}, increasing the number of the eavesdroppers can have more significant impact on the revenue than increasing the number of the legitimate users.
In Fig. \ref{fig:U_L_vs_No_PB_1}, the revenue is decreased when
the distance between the source and PBs is increased from 5m to 6.5m. This is because the nearer the PBs to the transmitter, the higher the transferred energy efficiency between them, which reduces the transmitter's payments to the PBs for their wireless energy services.
Fig. \ref{fig:Theta_vs_No_PB} shows the optimal energy transfer time $ \theta $ versus the number of PBs. It is observed that   $ \theta $ decreases as either  the number of the PBs or $ \mu $ increase. The same behavior is observed when the distance between the source and PBs (i.e., $ d_{PB} $) decreases.
Fig. \ref{fig:Optimal_price_vs_No_PB} shows the optimal energy price versus the number of PBs. The price decreases as the number of PBs increases. Besides, the larger $ \mu $, the higher optimal energy price needs to be paid. It can also be seen from this figure that the decrease of the distance between the transmitter and PBs can also reduce the optimal energy price. This is because the shorter the distance between the source and PBs, the more energy harvested by the transmitter for the same power transmitted by the PBs, such that a lower energy price can be paid by the transmitter. 

\section{Conclusion and Future Work}\label{section Conclusion}
In this paper, we investigated the secure wireless-powered multiantenna multicasting system, which consists of a dedicated wireless powered networks with multiple PBs and a multiantenna secure multicasting system. We first investigated the power minimization problem to obtain the optimal power and energy transfer time allocation when both wireless power and secure information networks belong to the same service supplier. As this problem is not convex, we proposed a two-level approach, where the inner level problem can be relaxed as a convex optimization framework via iterative conic convex reformulation (i.e., SCA based SOCP scheme), while the outer level problem can be handled by using an 1D search. Next, we formulate a \emph{Stackelberg} game for the case when the WET and secure WIT networks belong to different service suppliers. We formulated the \textit{energy trading} process between the transmitter and the PBs as a \emph{Stackelberg} game, in which the transmitter plays a leader role and pays a price for the energy services from the PBs to guarantee the required security, and optimizes the energy price and the energy transfer time to maximize its utility function. Meanwhile, the PBs are modelled as the followers that determine their optimal transmit powers based on the released energy price to maximize their own utility function. The \emph{Stackelberg} equilibrium have been derived in terms of closed-form solution, where both the transmitter and the PBs come to an agreement on the energy price, transmit power and energy transfer time. Simulation results have been provided to validate the proposed schemes. For future works, we can consider 
%the multiantenna multicasting system which is wireless powered by multiple PBs, and the PBs only work during the WET phase to provide the power for the transmitter. However, it is not always possible that the secrecy communication can be guaranteed by only relying on the WET from the PBs. Thus, we can consider 
a more challenging scenario such that the PBs will not stay silent during the WIT phase but rather transmit artificial noise to interfere with the eavesdroppers. This would change the dynamic of the optimization problems and may require different design/solutions.
%This case is similarly considered as two different scenarios, both WET and WIT networks belong to the same/different service providers. 
%The associated convex optimization and game theory techniques can then be employed to solve the formulated problem, and the optimal power allocation of the PBs will be analysed during WET and WIT phases. These approaches will be considered in our future work.

\begin{appendix}
	\subsection{Proof of \emph{Theorem} \ref{theorem power efficient schur complement}}\label{appendix therorem schur complement}
	When the solution $ \mathbf{Q}_{s} $ satisfies rank-one conditions in \emph{Lemma} \ref{lemma:Rank_one_lemma}, the SDP relaxed problem \eqref{eq:SDP_relaxation_results} can be equivalently modified as
		\begin{align}\label{eq:SOCP_power_efficient_problem_rewritten}
		\min_{\mathbf{w}}&~ \|\mathbf{w}\|^{2}, \nonumber\\
		s.t.&~ \frac{2^{\bar{\bar{R}}}}{\bar{\sigma_{e}^2}} |\mathbf{w}^{H}\mathbf{h}_{e,l}|^2 + (2^{\bar{\bar{R}}} - 1) \leq  \frac{1}{\bar{\sigma}_{s}^{2}} |\mathbf{w}^{H}\mathbf{h}_{s,k}|^{2}, ~\forall (k,l).
		\end{align} 
%	\begin{subequations}
%	\begin{align}
%\min_{t,\mathbf{w}} &~ t, \nonumber\\
%s.t. &~\|\mathbf{w}\|^2 \leq t, \label{eq:Power_constraint}\\
%&~\frac{2^{\bar{\bar{R}}}}{\bar{\sigma_{e}^2}} |\mathbf{w}^{H}\mathbf{h}_{e,l}|^2 + (2^{\bar{\bar{R}}} - 1) \leq \frac{1}{\bar{\sigma}_{s}^2} |\mathbf{w}^{H}\mathbf{h}_{s,k}|^2,~ \forall k,\forall l. \label{eq:Sec_rate_constraint}
%	\end{align}
%	\end{subequations}
	The problem \eqref{eq:SOCP_power_efficient_problem_rewritten} is still not convex in terms of the beamforming vector $ \mathbf{w} $. Thus, we propose a novel reformulation method by applying the following inequality relations in \eqref{eq:SOCP_power_efficient_problem_rewritten},
	\begin{align}\label{eq:SOC_theorem}
	\left[\begin{array}{cc}
	a \\ \mathbf{b}
	\end{array}
	\right] \succeq_{K_{n}} \mathbf{0} \Leftrightarrow \|\mathbf{b}\| \leq a,
	\end{align}
	such that \eqref{eq:SOCP_power_efficient_problem_rewritten} can be reformulated into the following SOCP: 
	\begin{align}
	\min_{t,\mathbf{w}} &~~ t, \nonumber\\
	s.t. &~ \left[\begin{array}{cc}
	t_{1} \\ \mathbf{w}
	\end{array} 
	\right] \succeq_{K_{N_{T}+1}} \mathbf{0}, ~ \left[\begin{array}{cc}
	\frac{1}{\bar{\sigma}_{s}} \mathbf{w}^{H}\mathbf{h}_{s,k}
	\\\frac{2^{\bar{\bar{R}}}}{\bar{\sigma}_{e}}\mathbf{w}^{H}\mathbf{h}_{e,l} \\
	(2^{\bar{\bar{R}}} - 1)^{\frac{1}{2}}
	\end{array}
	\right] \succeq_{K_{3}} \mathbf{0}, ~\forall k, \forall l.
	\end{align}
	Thus, this completes the proof of \emph{Theorem} \ref{theorem power efficient schur complement}. 
			\subsection{Proof of \emph{Lemma} \ref{lemma Special_case_power_efficient}}\label{appendix Special_case_power_efficient}
			The power minimization problem \eqref{eq:Power_efficient_for_w} with only single legitimate user and eavesdropper can be rewritten as 
			\begin{align}\label{eq:Power_efficient_for_special_case}
			\min_{\mathbf{w}} &~ \mathbf{w}^{H}\mathbf{w} \nonumber\\
			s.t. &~ \log\left(1+\frac{ |\mathbf{h}_{s}^{H}\mathbf{w}|^{2}}{\bar{\sigma}_{s}^{2}}\right) - \log\left(1+\frac{ |\mathbf{h}_{e}^{H}\mathbf{w}|^{2}}{\bar{\sigma}_{e}^{2}}\right) \geq \bar{\bar{R}}.
			\end{align}		
%			Setting $ \mathbf{w} = \sqrt{p}\mathbf{\tilde{w}} $, we have 
           Now we equivalently modify \eqref{eq:Power_efficient_for_special_case} as 
			\begin{align}\label{eq:Power_efficient_for_p}
			\min_{\mathbf{v},P} &~ P\mathbf{v}^{H}\mathbf{v} \nonumber\\
			s.t. &~ \frac{\mathbf{v}^{H}(\mathbf{I}+\frac{P}{\bar{\sigma}_{s}^{2}} \mathbf{h}_{s}\mathbf{h}_{s}^{H})\mathbf{v}}{\mathbf{v}^{H}(\mathbf{I}+\frac{P}{\bar{\sigma}_{e}^{2}} \mathbf{h}_{e}\mathbf{h}_{e}^{H})\mathbf{v}} \geq 2^{\bar{\bar{R}}},~ \mathbf{v}^{H}\mathbf{v} = 1,~ P \geq 0.
			\end{align}
%			The problem \eqref{eq:Power_efficient_for_p} consists two variables ($ p, \mathbf{\tilde{w}} $), 
			In order to achieve the optimal solution $ (P^{\textrm{opt}},\mathbf{v}^{\textrm{opt}}) $, we consider the Lagrange dual problem to \eqref{eq:Power_efficient_for_special_case}, which can be expressed as 
			\begin{align}
			\mathcal{L}(\mathbf{w},\mu) = \mathbf{w}^{H}\mathbf{w} + \alpha \left[ 2^{\bar{\bar{R}}}\bigg(1+\frac{1}{\bar{\sigma}_{s}^{2}}\mathbf{w}^{H}\mathbf{h}_{e}\mathbf{h}_{e}^{H}\mathbf{w}\bigg) - \bigg(1+ \frac{1}{\bar{\sigma}_{e}^{2}}\mathbf{w}^{H}\mathbf{h}_{s}\mathbf{h}_{s}^{H}\mathbf{w}\bigg)  \right],
			\end{align}
			where $ \alpha \geq 0 $ is the Lagrange multiplier associated with the secrecy rate constraint. The corresponding dual problem can be given by 
			\begin{align}\label{eq:Dual_problem_power_efficient}
			\max_{\alpha \geq 0} ~ \alpha(2^{\bar{\bar{R}}} - 1), ~s.t. ~ \mathbf{Y} = \mathbf{I} - \alpha \bigg(\frac{1}{\bar{\sigma}_{s}^{2}}\mathbf{h}_{s}\mathbf{h}_{s}^{H} - \frac{1}{\bar{\sigma}_{e}^{2}}\mathbf{h}_{e}\mathbf{h}_{e}^{H}\bigg) \succeq \mathbf{0}. 
			\end{align}
			The constraint in \eqref{eq:Dual_problem_power_efficient} implies that the matrix $ \mathbf{Y} $ have at least one zero eigenvalue. On the other hand, the solution of $ \alpha $ can be the maximum value that satisfies the positive semidefinite constraint in \eqref{eq:Dual_problem_power_efficient}, which leads to 
			\begin{align}
			\alpha^{\textrm{opt}} = \frac{1}{\varrho_{\textrm{max}} (\frac{1}{\bar{\sigma}_{s}^{2}}\mathbf{h}_{s}\mathbf{h}_{s}^{H}-\frac{2^{\bar{\bar{R}}}}{\bar{\sigma}_{e}^{2}}\mathbf{h}_{e}\mathbf{h}_{e}^{H})}.
			\end{align}
			The problem \eqref{eq:Power_efficient_for_special_case} can be formulated as a convex optimization problem. Hence, the strong duality holds between the original problem \eqref{eq:Power_efficient_for_special_case} and the corresponding dual problem \eqref{eq:Dual_problem_power_efficient}. The required minimum power to achieve the secrecy rate constraint is 
			\begin{align}
			P^{\textrm{opt}} = \alpha^{\textrm{opt}}(2^{\bar{\bar{R}}} - 1).
			\end{align}
			On the other hand, it is easily verified that the optimal $ \mathbf{w} $ lies in the null space of $ \mathbf{Y} $
			\begin{align}
			\mathbf{\bar{w}} = v_{\textrm{max}}
			\bigg(\frac{1}{\bar{\sigma}_{s}^{2}}\mathbf{h}_{s}\mathbf{h}_{s}^{H} - 
			\frac{2^{\bar{\bar{R}}}}{\bar{\sigma}_{e}^{2}}\mathbf{h}_{e}\mathbf{h}_{e}^{H}\bigg),~ \mathbf{v}^{\textrm{opt}} = \frac{\mathbf{\bar{w}}}{\|\mathbf{\bar{w}}\|_{2}}. 
			\end{align}
			Hence, the optimal solution to \eqref{eq:Power_efficient_for_special_case} can be expressed as 
			\begin{align}
			\mathbf{w}^{\textrm{opt}} = \sqrt{P^{\textrm{opt}}} \mathbf{v}^{\textrm{opt}}.
			\end{align}
%			This completes the proof of \emph{Lemma} \ref{lemma Special_case_power_efficient}.
			\subsection{Proof of \emph{Lemma} \ref{lemma Second_step_convex_show}}\label{appendix Second_step_convex_show}
			In order to show that $f(\theta)$ is a convex problem, there are two conditions to be satisfied: convex objective function and convex constraints \cite{boyd_B04}. First, we rewrite $f(\theta)$ as follows:
			\begin{align}\label{eq:f_theta}
		 f(\theta)  = \min_{\theta} &~ \frac{\theta(\sum_{m=1}^{M} \|\mathbf{g}_{m}\|^{2})}{1-\theta} \|\mathbf{w}\|^{2}, \nonumber\\
			s.t. &~ (1-\theta) \log\left(\frac{1-\theta + \theta t_{s,k}}{1-\theta + \theta t_{e,l}}\right) \geq \bar{R},
			0 < \theta < 1,~\forall k,\forall l.
			\end{align}		
			It is easily verified that the objective function in \eqref{eq:f_theta} is convex with respect to $ \theta $. Then, we show that the secrecy constraint in \eqref{eq:f_theta} is convex. 
			Let $ f_{k,l}(\theta) = (1-\theta) \log\left(\frac{1-\theta + \theta t_{s,k}}{1-\theta + \theta t_{e,l}}\right) $, where $ t_{s,k} = \frac{|\mathbf{h}_{s,k}^{H}\mathbf{w}|^{2}}{\sigma_{s}^{2}} $ and $ t_{e,l} = \frac{|\mathbf{h}_{e,l}^{H}\mathbf{w}|^{2}}{\sigma_{e}^{2}} $. Hence, the remaining part we only show that $ f_{k,l}(\theta) $ is a concave function with respect to $ \theta $, which can be equivalently written as 
			\begin{align}
			f_{k,l}(\theta) &~ = (1-\theta) \log\left( \frac{\frac{1-\theta+\theta t_{s,k}}{1-\theta}}{\frac{1-\theta +\theta t_{e,l}}{1-\theta}} \right) = (1-\theta) \log\left( \frac{\frac{\theta(t_{s,k} -1)+(1-t_{s,k}) +t_{s,k}}{1-\theta}}{\frac{\theta(t_{e,l} - 1) + (1-t_{e,l}) + t_{e,l}}{1-\theta}} \right) \nonumber\\
			&~ = \log\left( \frac{\frac{(1-t_{s,k})(1-\theta) + t_{s,k}}{1-\theta}}{\frac{(1-t_{e,l})(1-\theta)+t_{e,l}}{1-\theta}} \right). 
			\end{align}
			Let $ z = 1- \theta $, ($ 0\leq \theta \leq 1 $), $ f_{k,l}(z) $ can be rewritten as 
			\begin{align}
			f_{k,l}(z) = z \log \frac{(1-t_{s,k})z +t_{s,k}}{z} - z \log \frac{(1-t_{e,l})z +t_{e,l}}{z}.
			\end{align}
			Then, we consider the first derivative of $ f_{k,l}(z) $, 
			\begin{align}
			\frac{ \partial f_{k,l}(z)}{z} = \frac{1}{\ln2 } \left[ \bigg( \ln \frac{(1-t_{s,k})z+t_{s,k}}{z} + \frac{-t_{s,k}}{(1-t_{s,k})z + t_{s,k}}\bigg) - \bigg( \ln \frac{(1-t_{e,l})z+t_{e,l}}{z} + \frac{-t_{e,l}}{(1-t_{e,l})z + t_{e,l}}\bigg) \right].
			\end{align}
			Furthermore, the second derivative of $ f_{k,l}(z) $ is given by 
			\begin{align}
			\frac{\partial^{2} f_{k,l}(z)}{z^{2}} = \frac{1}{\ln2} \left[ \frac{-t_{s,k}^{2}}{[(1-t_{s,k})z+t_{s,k}]^{2} z} - \frac{-t_{e,l}^{2}}{[(1-t_{e,l})z +t_{e,l}]^{2} z} \right].
			\end{align}
			Let $ g(t) = \frac{-t^{2}}{[(1-t)z+t]^{2}z} $, the first derivative of $ g(t) $ is given by 
			\begin{align}\label{eq:g(t)_first_order_derivatives}
			\frac{\partial g(t)}{t} &~ = \frac{-2t[(1-t)z + t] + 2(1-z)t^{2}}{[(1-t)z + t]^{3}z} = \frac{-2zt}{[(1-t)z+t]^{3}z} < 0. 
			\end{align}
			It is easily verified that \eqref{eq:g(t)_first_order_derivatives} holds since $ z = 1- \theta \in (0,1) $, thus, $ g(t) $ is a monotonically decreasing function of $ t $. Due to $ t_{s,k} > t_{e,l} $, it is easily obtained that $ \frac{\partial^{2} f_{k,l}(z)}{z^{2}} < 0 $. In other words, $ f_{k,l}(\theta) $ is a concave function with respect to $ \theta $. 
			\subsection{Proof of \emph{Lemma} \ref{lemma lambda_concave_function}}\label{appendix lambda_concave_function}
				We first derive the first-order derivatives of \eqref{eq:U_L_lambda}, which is written as
				\begin{align}
				\frac{\partial U_{M}(\lambda)}{\partial \lambda} = \frac{1}{\ln 2}\bigg[ \frac{\mu (1-\theta) t_{s} C_{M} }{1+(\lambda C_{M} - 2 D_{M}) t_{s}} - \frac{\mu (1-\theta) t_{e} C_{M} }{1+(\lambda C_{M} -2 D_{M}) t_{e}} \bigg] - 2 C_{M} \lambda + 2 D_{M}. 
				\end{align}	
				Then, the second-order derivatives of \eqref{eq:U_L_lambda} is given by
				\begin{align}
				\frac{\partial^{2} U_{M}(\lambda)}{\partial^{2} \lambda} = \frac{1}{\ln 2}\bigg[ \frac{\mu (1-\theta) C_{M}^{2}(t_{e}^{2} - t_{s}^{2})}{[1-(\lambda C_{M} - 2 D_{M})t_{s}]^{2} [1-(\lambda C_{M} - 2 D_{M})t_{e}]^{2}} \bigg] - 2C_{M}  < 0.
				\end{align}
				The above inequality holds since $ t_{s} - t_{e} > 0 $ to guarantee the minimum achievable secrecy rate is greater than zero. Thus, \eqref{eq:U_L_lambda} is a concave function.
\end{appendix}

\bibliographystyle{ieeetr}
\bibliography{my_references}

\begin{thebibliography}{10}

\bibitem{Tom_Luo_Multicasting_TSP_2006}
N.~D. Sidiropoulos, T.~N. Davidson, and Z.-Q. Luo, ``Transmit beamforming for
  physical-layer multicasting,'' {\em IEEE Transactions on Signal Processing},
  vol.~54, no.~6, pp.~2239--2251, Jun.~2006.

\bibitem{Zhengzheng_Xiang_TWC_Multicasting_2013}
Z.~Xiang, M.~Tao, and X.~Wang, ``Coordinated multicast beamforming in multicell
  networks,'' {\em IEEE Trans. Wireless Commun.}, vol.~12, no.~1, pp.~12--21,
  Jan.~2013.

\bibitem{Zhengzheng_Xiang_JSAC_Multicasting_2014}
Z.~Xiang, M.~Tao, and X.~Wang, ``Massive mimo multicasting in noncooperative
  cellular networks,'' {\em IEEE J. Sel. Area. Comm.}, vol.~32, no.~6,
  pp.~1180--1193, Jun.~2014.

\bibitem{Varshney_08}
L.~Varshney, ``Transporting information and energy simultaneously,'' in {\em
  Proc. 2008 IEEE Int. Symp. Inf. Theory}, pp.~1612--1616, July, 2008.

\bibitem{Shannon_Tesla_10}
P.~Grover and A.~Sahai, ``Shannon meets tesla: Wireless information and power
  transfer,'' in {\em Proc. 2010 IEEE Int. Symp. Inf. Theory}, pp.~2363--2367,
  Jun., 2010.

\bibitem{Zhang_Rui_SWIPT_TWC13}
R.~Zhang and C.~K. Ho, ``{MIMO} broadcasting for simultaneous wireless
  information and power transfer,'' {\em IEEE Trans.~Wireless Commun.},
  vol.~12, pp.~1989--2001, May 2013.

\bibitem{Rui_Zhang_WPCN_CM_2016}
S.~Bi, Y.~Zeng, and R.~Zhang, ``Wireless powered communication networks: an
  overview,'' {\em IEEE Wireless Commun.}, vol.~23, no.~2, pp.~10--18,
  Apr.~2016.

\bibitem{Rui_Zhang_WPCN_TWC_2014}
H.~Ju and R.~Zhang, ``Throughput maximization in wireless powered communication
  networks,'' {\em IEEE Trans.~Wireless Commun.}, vol.~13, no.~1, pp.~418--428,
  Jan.~2014.

\bibitem{Rui_Zhang_GLOBECOM_2014_UC_WPCN}
H.~Ju and R.~Zhang, ``User cooperation in wireless powered communication
  networks,'' in {\em Proc. IEEE GLOBECOM}, pp.~1430--1435, Dec. 2014.

\bibitem{He_Chen_ITW_WPCN_Relay_2014}
H.~Chen, X.~Zhou, Y.~Li, P.~Wang, and B.~Vucetic, ``Wireless-powered
  cooperative communications via a hybrid relay,'' in {\em Proc. IEEE
  Information Theory Workshop (ITW)}, pp.~666--670, Nov. 2014.

\bibitem{He_Chen_TSP_HTC_WPCN_2016}
H.~Chen, Y.~Li, J.~L. Rebelatto, B.~F. Uchôa-Filho, and B.~Vucetic,
  ``Harvest-then-cooperate: Wireless-powered cooperative communications,'' {\em
  IEEE Trans.~Signal Process.}, vol.~63, no.~7, pp.~1700--1711, Apr.~2015.

\bibitem{Kaibin_Huang_CM_2015}
K.~Huang and X.~Zhou, ``Cutting the last wires for mobile communications by
  microwave power transfer,'' {\em IEEE Commun. Mag.}, vol.~53, no.~6,
  pp.~86--93, Jun.~2015.

\bibitem{Kaibin_Huang_TWC_2014_WPT}
K.~Huang and V.~K.~N. Lau, ``Enabling wireless power transfer in cellular
  networks: Architecture, modeling and deployment,'' {\em IEEE Trans.~Wireless
  Commun.}, vol.~13, no.~2, pp.~902--912, Feb.~2014.

\bibitem{Caijun_Zhong_SPL_2015_WPR}
C.~Zhong, G.~Zheng, Z.~Zhang, and G.~K. Karagiannidis, ``Optimum wirelessly
  powered relaying,'' {\em IEEE Signal Process. Lett.}, vol.~22, no.~10,
  pp.~1728--1732, Oct.~2015.

\bibitem{Wyner_J75}
A.~D. Wyner, ``The wire-tap channel,'' {\em {Bell Syst. Tech. Journ.}},
  vol.~54, pp.~1355--1387, Jan. 1975.

\bibitem{Korner_Info_Theory_J78}
{I.~Csisz\'{a}r and J.~K\"{o}rner}, ``Broadcast channels with confidential
  messages,'' {\em IEEE Trans. Inform. Theory}, vol.~24, pp.~339--348, May
  1978.

\bibitem{Shamai_Multiantenna_Wiretap_TIT_2009}
T.~Liu and S.~Shamai, ``A note on the secrecy capacity of the multiple-antenna
  wiretap channel,'' {\em IEEE Trans.~Inform. Theory}, vol.~55, no.~6,
  pp.~2547--2553, Jun.~2009.

\bibitem{Wornell_Info_Theory_J10}
A.~Khisti and G.~W. Wornell, ``{Secure transmission with multiple antennas I:
  The MISOME wiretap channel},'' {\em IEEE Trans. Inform. Theory}, vol.~56,
  no.~7, pp.~3088--3104, Jul.~2010.

\bibitem{Wornell_Info_Theory1_J10}
A.~Khisti and G.~W. Wornell, ``{Secure transmission with multiple antennas II:
  The MIMOME wiretap channel},'' {\em IEEE Trans. Inform. Theory}, vol.~56,
  no.~11, pp.~5515--5532, Nov.~2010.

\bibitem{Petropulu_Gaussian_MISO_Wiretap_TWC_2011}
J.~Li and A.~P. Petropulu, ``On ergodic secrecy rate for gaussian {MISO}
  wiretap channels,'' {\em IEEE Trans.~Wireless Commun.}, vol.~10,
  pp.~1176--1187, April Apr.~2011.

\bibitem{Swindlehurst_FR_MIMO_Wiretap_TSP_2013}
S.~A.~A. Fakoorian and A.~L. Swindlehurst, ``Full rank solutions for the {MIMO}
  gaussian wiretap channel with an average power constraint,'' {\em IEEE Trans
  on Signal Process.}, vol.~61, no.~10, pp.~2620--2631, May.~2013.

\bibitem{Ma_Sig_Process_J11}
Q.~Li and W.-K. Ma, ``{Optimal and robust transmit designs for MISO channel
  secrecy by semidefinite programming},'' {\em IEEE Trans. Signal Process.},
  vol.~59, no.~8, pp.~3799--3812, Aug. 2011.

\bibitem{Ma_TSP_J13}
Q.~Li and W.-K. Ma, ``Spatially selective artificial-noise aided transmit
  optimization for {MISO} multi-eves secrecy rate maximization,'' {\em IEEE
  Trans. Signal Process.}, vol.~61, no.~10, pp.~2704--2717, May~2013.

\bibitem{Zheng_Secrecy_J15}
Z.~Chu, K.~Cumanan, Z.~Ding, M.~Johnston, and S.~Le~Goff, ``Secrecy rate
  optimizations for a {MIMO} secrecy channel with a cooperative jammer,'' {\em
  IEEE Trans.~Vehicular Technol.}, vol.~64, no.~5, pp.~1833--1847, May~2015.

\bibitem{Zheng_Stackelberg_game_EUSIPCO_2014}
Z.~Chu, K.~Cumanan, Z.~Ding, M.~Johnston, and S.~L. Goff, ``Secrecy rate
  optimization for a {MIMO} secrecy channel based on stackelberg game,'' in
  {\em Proc.~European Signal Processing Conference (EUSIPCO), Lisbon,
  Portugal}, pp.~126--130, Sept.~2014.

\bibitem{Zheng_WCL_2015}
Z.~Chu, K.~Cumanan, Z.~Ding, M.~Johnston, and S.~Le~Goff, ``Robust outage
  secrecy rate optimizations for a {MIMO} secrecy channel,'' {\em
  IEEE,~Wireless Commun. Lett.}, vol.~4, no.~1, pp.~86--89, Feb.~2015.

\bibitem{Zheng_Sec_TWC_2016}
Z.~Chu, H.~Xing, M.~Johnston, and S.~L. Goff, ``Secrecy rate optimizations for
  a {MISO} secrecy channel with multiple multiantenna eavesdroppers,'' {\em
  IEEE Trans.~Wireles Commun.}, vol.~15, no.~1, pp.~283--297, Jan.~2016.

\bibitem{Nallan_TCOM_2015_CCR_Sec_Game}
A.~Al-Talabani, Y.~Deng, A.~Nallanathan, and H.~X. Nguyen, ``Enhancing secrecy
  rate in cognitive radio networks via stackelberg game,'' {\em IEEE
  Trans.~Commun.}, vol.~64, no.~11, pp.~4764--4775, Nov.~2016.

\bibitem{Huiming_Wang_RFID_Backscatter_PLS_TWC_2016}
Q.~Yang, H.~M. Wang, Y.~Zhang, and Z.~Han, ``Physical layer security in {MIMO}
  backscatter wireless systems,'' {\em IEEE Trans.~Wireless Commun.}, vol.~15,
  no.~11, pp.~7547--7560, Nov.~2016.

\bibitem{Zhang_Rui_TSP_Sec_SWIPT_J14}
L.~Liu, R.~Zhang, and K.-C. Chua, ``Secrecy wireless information and power
  transfer with {MISO} beamforming,'' {\em IEEE Trans. Signal Process.},
  vol.~62, no.~7, pp.~1850--1863, Apr. ~2014.

\bibitem{Derrick_SWIPT_TWC_2014}
D.~W.~K. Ng, E.~S. Lo, and R.~Schober, ``Robust beamforming for secure
  communication in systems with wireless information and power transfer,'' {\em
  IEEE Trans.~Wireless Commun.}, vol.~13, no.~8, pp.~4599--4615, Aug.~2014.

\bibitem{Khandaker_TIFS_J15}
M.~Khandaker and K.~Wong, ``Masked beamforming in the presence of
  energy-harvesting eavesdroppers,'' {\em IEEE Trans.~Inf. Forensics Security},
  vol.~10, pp.~40--54, Jan 2015.

\bibitem{Zheng_TVT_SWIPT_2015}
Z.~Chu, Z.~Zhu, M.~Johnston, and S.~Y.~L. Goff, ``Simultaneous wireless
  information power transfer for {MISO} secrecy channel,'' {\em IEEE
  Trans.~Vehicular Technol.}, vol.~65, no.~9, pp.~6913--6925, Sept.~2016.

\bibitem{Qiang_Li_ICC_Sec_Multicasting_2011}
Q.~Li and W.~K. Ma, ``Multicast secrecy rate maximization for {MISO} channels
  with multiple multi-antenna eavesdroppers,'' in {\em Proc. IEEE,~ICC, Tokyo,
  Japan}, pp.~1--5, Jun. 2011.

\bibitem{Cumanan_JSTSP_2016_Game}
K.~Cumanan, Z.~Ding, M.~Xu, and H.~V. Poor, ``Secrecy rate optimization for
  secure multicast communications,'' {\em IEEE J. Sel. Topics Signal Process.},
  vol.~10, no.~8, pp.~1417--1432, Dec.~2016.

\bibitem{Ulukus_ISIT_2007_MISO_Sec}
S.~Shafiee and S.~Ulukus, ``Achievable rates in gaussian {MISO} channels with
  secrecy constraints,'' in {\em Proc.~IEEE ISIT,~Nice,~France},
  pp.~2466--2470, Jun.~2007.

\bibitem{Xiaoming_Chen_CL_2016_Sec_WPCN}
Y.~Wu, X.~Chen, C.~Yuen, and C.~Zhong, ``Robust resource allocation for secrecy
  wireless powered communication networks,'' {\em IEEE Commun. Lett.}, vol.~20,
  no.~12, pp.~2430--2433, Dec.~2016.

\bibitem{Mukherjee_ICASSP_2012}
A.~Mukherjee and A.~L. Swindlehurst, ``Detecting passive eavesdroppers in the
  {MIMO} wiretap channel,'' in {\em IEEE ICASSP,~Tokyo,~Japan}, Mar.~2012.

\bibitem{Geraci_TCOM_2012}
G.~Geraci, M.~Egan, J.~Yuan, A.~Razi, and I.~B. Collings, ``Secrecy sum-rates
  for multi-user {MIMO} regularized channel inversion precoding,'' {\em IEEE
  Trans.~Commun.}, vol.~60, no.~11, pp.~3472--3482, Nov.~2012.

\bibitem{boyd_B04}
S.~Boyd and L.~Vandenberghe, {\em Convex Optimization}.
\newblock Cambridge, UK: Cambridge University Press, 2004.

\bibitem{Yongwei_Huang_Rank_Reduction_TSP_2010}
Y.~Huang and D.~P. Palomar, ``Rank-constrained separable semidefinite
  programming with applications to optimal beamforming,'' {\em IEEE
  Trans.~Signal Process.}, vol.~58, no.~2, pp.~664--678, Feb.~2010.

\bibitem{Kunyu_Wang_TSP_2014_Outage}
K.~Y. Wang, A.~M.~C. So, T.~H. Chang, W.~K. Ma, and C.~Y. Chi, ``Outage
  constrained robust transmit optimization for multiuser {MISO} downlinks:
  Tractable approximations by conic optimization,'' {\em IEEE Trans.~Signal
  Process.}, vol.~62, no.~21, pp.~5690--5705, Nov.~2014.

\bibitem{Barry_Marks_Approx_Noncvx_Math_Programs_OR_1978}
B.~R. Marks and G.~P. Wright, ``A general inner approximation algorithm for
  nonconvex mathematical programs,'' {\em Oper Res.}, vol.~26, no.~4,
  pp.~681--683, 1978.

\bibitem{LN_Tran_SPL_Fast_converge_2012}
L.~N. Tran, M.~F. Hanif, A.~Tolli, and M.~Juntti, ``Fast converging algorithm
  for weighted sum rate maximization in multicell {MISO} downlink,'' {\em IEEE
  Signal Process. Lett.}, vol.~19, no.~12, pp.~872--875, Dec.~2012.

\bibitem{LN_Tran_SPL_Large_Scale_2014}
L.~N. Tran, M.~F. Hanif, and M.~Juntti, ``A conic quadratic programming
  approach to physical layer multicasting for large-scale antenna arrays,''
  {\em IEEE Signal Process. Lett.}, vol.~21, no.~1, pp.~114--117, Jan.~2014.

\bibitem{Schober_TSG2010_DSM_GT}
A.~H. Mohsenian-Rad, V.~W.~S. Wong, J.~Jatskevich, R.~Schober, and
  A.~Leon-Garcia, ``Autonomous demand-side management based on game-theoretic
  energy consumption scheduling for the future smart grid,'' {\em IEEE
  Trans.~Smart Grid}, vol.~1, no.~3, pp.~320--331, Dec.~2010.

\bibitem{Xiaobin_Huang_CL_2014_FD}
X.~Huang, J.~He, Q.~Li, Q.~Zhang, and J.~Qin, ``Optimal power allocation for
  multicarrier secure communications in full-duplex decode-and-forward relay
  networks,'' {\em IEEE Commun. Lett.}, vol.~18, no.~12, pp.~2169--2172,
  Dec.~2014.

\bibitem{Jialing_Liao_CL_Robust_PS_SWIPT_2016}
J.~Liao, M.~R.~A. Khandaker, and K.~K. Wong, ``Robust power-splitting {SWIPT}
  beamforming for broadcast channels,'' {\em IEEE Commun. Lett.}, vol.~20,
  no.~1, pp.~181--184, Jan.~2016.

\end{thebibliography}

\end{document}